\documentclass[twocolumn]{aastex631}
\usepackage{printlen}
\usepackage{placeins}

\begin{document}

\title{ExoGemS High-Resolution Transmission Spectroscopy of WASP-76b with GRACES}

\author[0000-0001-9796-2158]{Emily K.~Deibert}
\affiliation{Gemini Observatory, NSF’s NOIRLab, Casilla 603, La Serena, Chile}

\author[0000-0001-6391-9266]{Ernst J. W. de Mooij}
\affil{Astrophysics Research Centre, Queen's University Belfast, Belfast BT7 1NN, UK}

\author[0000-0001-5349-6853]{Ray Jayawardhana}
\affil{Department of Astronomy, Cornell University, Ithaca, New York 14853, USA}

\author[0000-0001-7836-1787]{Jake D. Turner}
\affil{Department of Astronomy, Cornell University, Ithaca, New York 14853, USA}
\affil{NHFP Sagan Fellow}

\author[0000-0002-5425-2655]{Andrew Ridden-Harper}
\affil{Las Cumbres Observatory, 6740 Cortona Drive, Suite 102, Goleta, CA 93117, USA}

\author[0000-0003-1150-7889]{Callie E. Hood}
\affil{Department of Astronomy \& Astrophysics, University of California, Santa Cruz, CA 95064, USA}

\author[0000-0002-9843-4354]{Jonathan J. Fortney}
\affil{Department of Astronomy \& Astrophysics, University of California, Santa Cruz, CA 95064, USA}

\author[0000-0001-6362-0571]{Laura Flagg}
\affil{Department of Astronomy, Cornell University, Ithaca, New York 14853, USA}

\author[0000-0003-4426-9530]{Luca Fossati}
\affil{Space Research Institute, Austrian Academy of Sciences, Schmiedlstrasse 6, A-8042 Graz, Austria}

\author[0000-0002-1199-9759]{Romain Allart}
\affil{Department of Physics, and Institute for Research on Exoplanets, Universit\'e de Montr\'eal, Montr\'eal, H3T 1J4, Canada}

\author[0000-0002-7704-0153]{Matteo Brogi}
\affil{Dipartimento di Fisica, Universit\`a degli Studi di Torino, via Pietro Giuria 1, I-10125, Torino, Italy}
\affil{INAF-Osservatorio Astrofisico di Torino, Via Osservatorio 20,
I-10025 Pino Torinese, Italy}

\author[0000-0003-4816-3469]{Ryan J. MacDonald}
\affil{Department of Astronomy, University of Michigan, 1085 S. University Ave., Ann Arbor, MI 48109, USA}
\affil{NHFP Sagan Fellow}
 
\begin{abstract}
We present high-resolution transmission spectroscopy of WASP-76b with GRACES/Gemini North obtained as part of the ExoGemS survey. With a broad spectral range of 400--1050 nm and a relatively high resolution of $\sim$66,000, these observations are particularly well-suited to searching for atomic and molecular atmospheric species via the Doppler cross-correlation technique.
We recover absorption features due to neutral iron (Fe~I), sodium (Na~I), and ionized calcium (Ca~II) at high significance ($>$$5\sigma$), and investigate possible atmospheric temperatures and wind speeds.
We also report tentative ($>$$3\sigma$) detections of Li~I, K~I, Cr~I, and V~I in the atmosphere of WASP-76b.
Finally, we report non-detections of a number of other species, some of which have previously been detected with other instruments.
Through model injection/recovery tests, we demonstrate that many of these species are not expected to be detected in our observations.
These results allow us to place GRACES and the ExoGemS survey in context with other high-resolution optical spectrographs.
\end{abstract}


\section{Introduction} \label{sec:intro}
In the two decades since the first detection of an exoplanet's atmosphere \citep{Charbonneau02}, the field of atmospheric characterization has flourished. Exoplanet atmospheres are now regularly observed from both the ground and space, and dozens of exoplanets have had their atmospheres detected and characterized to date. In recent years, high-resolution spectroscopy from ground-based instruments has been recognized as a particularly promising probe of exoplanet atmospheres at both optical and near-infrared wavelengths. High-resolution spectra allow us to resolve features from both atomic and molecular species, and the broad wavelength coverages of modern \'{E}chelle spectrographs allow us to detect many hundreds or even thousands of these features, boosting the strength of our detections \citep[see e.g.,][]{Birkby18} and allowing us to place robust constraints on the chemical compositions of a range of alien worlds.

With a relatively high resolving power (R $\sim$ 66,000) and broad wavelength coverage across the full optical range, the Gemini Remote Access to CFHT ESPaDOnS Spectrograph \citep[GRACES;][]{GRACES} at the Gemini North telescope is a robust tool for characterizing exoplanet atmospheres. The ongoing Gemini Large and Long Program GN-2020B-LP106: ``Exploring the Diversity of Exoplanet Atmospheres at High Spectral Resolution'' (Exoplanets with Gemini Spectroscopy or ExoGemS for short; PI: Jake Turner) aims to take advantage of these capabilities in order to carry out a systematic, high-resolution, comparative survey of transiting exoplanet atmospheres ranging from sub-Neptunes to ultra-hot Jupiters. The survey is expected to target a few dozen transiting exoplanets, many of which have not previously been observed at high spectral resolution. The goal of the survey is to compare the atmospheric properties and compositions of exoplanets across masses, temperatures, and stellar irradiation levels in order to determine the role that these properties play in regulating exoplanet atmospheres.

% WASP-76b
Owing to its high equilibrium temperature, short orbital period ($\sim$1.8 days; \citealt{Ehrenreich20}), and relatively bright host star (V~=~9.52; \citealt{SIMBAD}), WASP-76b \citep{West16} is an ideal benchmark target for the ExoGemS survey. We observed WASP-76b in the first semester of the survey (Gemini semester 2020B; see Section \ref{sec:obs} and Table \ref{tab:obs}) for the purposes of comparing the detection capabilities of GRACES with other high-resolution spectrographs (many of which have been used to observe WASP-76b) while simultaneously furthering our understanding of giant planet atmospheres by searching for atomic and molecular absorption features across the full optical range.

Since its discovery in 2016 \citep{West16}, WASP-76b has quickly become among the most well-studied ultra-hot Jupiters (i.e., hot Jupiters with equilibrium temperatures upwards of $\sim$2000~K; \citealt{Parmentier18, Bell18, Arcangeli18}). At these extreme temperatures, molecules are expected to dissociate and many atoms are expected to ionize, resulting in optical spectra rich in neutral and ionized atomic features that are amenable to detection via high-resolution spectroscopy \citep[e.g.,][]{Hoeijmakers19,BelloArufe22}.

Indeed, a wide variety of atomic features have been detected in the optical spectrum of WASP-76b. High-resolution detections of Na~I in its atmosphere were first reported by \cite{Seidel19} and \cite{Zak19} using HARPS/ESO 3.6m spectra. More recently, \cite{Ehrenreich20} reported a detection of asymmetric Fe~I absorption using observations from ESPRESSO/VLT, which was verified by \cite{Kesseli21} with HARPS and later investigated via modelling by \cite{Wardenier21} and \cite{Savel22}. Many other atomic species have also been reported in its atmosphere by \cite{Tabernero20}, \cite{Kesseli22}, and \cite{AS22} using the same ESPRESSO spectra analyzed in \cite{Ehrenreich20}. \cite{Seidel21} also used these spectra to provide updated measurements of Na~I absorption. Ca~II absorption in the form of the near-infrared triplet was also recently reported by \cite{CB21} with CARMENES/Calar Alto and \cite{Deibert21b}, the latter of which used the same GRACES spectra which are the focus of the present work. A summary of previous detections as of the writing of this paper is presented in Table \ref{tab:detections}. We note that a number of studies have also characterized WASP-76b's atmosphere in the near-infrared; however, we focus the present work on optical observations in order to place previous work into context with the ExoGemS survey.

\begin{deluxetable*}{ccccc}
\label{tab:detections}
\tablecaption{Summary of species detected in WASP-76b's atmosphere at high spectral resolution in the optical. Note that a ``detection''  or ``tentative detection'' may be defined differently in different works.}
\tablehead{%
    \colhead{Reference} & \colhead{Instrument/Telescope} & \colhead{Detected Species} & \colhead{Tentative} & \colhead{Upper Limits}
    }
\startdata
\cite{Zak19} & HARPS/ESO 3.6m & Na~I & & \\
\cite{Seidel19} & HARPS/ESO 3.6m & Na~I & & \\
\cite{Ehrenreich20} & ESPRESSO/VLT & Fe~I & & \\
\cite{Tabernero20} & ESPRESSO/VLT & Li~I, Na~I, Mg~I, Ca~II, & H$\alpha$\tablenotemark{a} & Ti~I, Cr~I, Ni~I, \\
 & & Mn~I, K~I, Fe~I & & TiO, VO, ZrO \\
\cite{Seidel21} & ESPRESSO/VLT; & Na~I & \\
 & HARPS/ESO 3.6m & & \\
\cite{Kesseli21} & HARPS/ESO 3.6m & Fe~I & & \\
\cite{CB21} & CARMENES/Calar Alto & Ca~II & & Li~I, K~I, H$\alpha$, Na~I \\
\cite{Deibert21b} & GRACES/Gemini North & Na~I, Ca~II & Li~I, K~I & H$\alpha$ \\
\cite{Kesseli22} & ESPRESSO/VLT & Li~I, Na~I, Mg~I, Ca~II, & H$\alpha$, K~I, Co~I & Ti~I, Ti~II, Ca~I, Al~I, \\
& & V~I, Cr~I, Mn~I, Fe~I, & & Sc~I, Sc~II, Zr~I\tablenotemark{b} \\
& & Ni~I, Sr~II & & \\
\cite{AS22} & ESPRESSO/VLT & Ba~II, Li~I, Na~I, Mg~I, & & \\
& & Ca~II, V~I, Cr~I, Mn~I, & & \\
& & Fe~I, H$\alpha$ & & \\
\enddata
\tablenotetext{a}{The authors detect H$\alpha$ at a significance of 4$\sigma$ in one transit, but are unable to recover the signal in the second transit.}
\tablenotetext{b}{The authors present 4$\sigma$ upper limits for a number of additional species which were not expected to be detectable in their analysis. For the sake of brevity, we have only displayed the 4$\sigma$ upper limits for species which the authors expected to be detectable.}
\end{deluxetable*}

In \cite{Deibert21b}, we demonstrated the efficacy of GRACES in characterizing exoplanet atmospheres via {single-line} transmission spectroscopy, allowing us to detect and resolve individual absorption lines in WASP-76b's atmosphere. We showed that GRACES is sensitive to detections of Na~I and the Ca~II infrared triplet, the latter of which may be an important probe of non-local thermodynamic equilibrium (NLTE) effects in exoplanet atmospheres \citep[e.g.,][]{Turner20, Fossati20, Fossati21, Deibert21b}. We also showed that GRACES is marginally sensitive to a number of other species using these methods, including Li~I, H$\alpha$, and K~I.

As in \cite{Deibert21b}, the goal of the present work is to search for species present in WASP-76b's atmosphere, allowing us to place GRACES into context among other high-resolution optical spectrographs and assess our detection capabilities for the remainder of the ExoGemS survey. 
Yet while \cite{Deibert21b} focused on single-line detections of strong absorbers such as Na~I and Ca~II via the creation of transmission spectra, in this work we turn our focus to detections of additional species via the Doppler cross-correlation technique \citep[e.g.,][]{Snellen10}, which combines the signals from weaker absorption lines which cannot be resolved individually. This technique is well-suited to detecting species with hundreds or even thousands of spectral features across a broad wavelength range. We thus focus our search on atoms and molecules with a large number of optical spectral features that can be resolved with GRACES.

This paper will proceed as follows. In Section \ref{sec:obs}, we describe the observations obtained as part of the ExoGemS survey with GRACES/Gemini North. We detail our data reduction methods in Section \ref{sec:reduc}, and in Section \ref{sec:analysis} we describe the methods we use to analyze the data, as well as the atmospheric models and templates used in this work. Our results are presented and discussed in Section \ref{sec:results}, and we conclude in Section \ref{sec:conclusion}.

\section{Observations}
\label{sec:obs}

We observed one transit of WASP-76b with GRACES \citep{GRACES} at the Gemini North telescope. The observations were obtained as part of the ExoGemS survey, which is a Gemini Large and Long Program to observe dozens of transiting exoplanets with high-resolution spectroscopy through the 2023A observing semester (GN-2020B-LP-106; PI: Jake Turner). A subset of these data was previously analyzed in \cite{Deibert21b}, where we derived transmission spectra around individual lines to search for atmospheric absorption, and reported detections of Ca~II and Na~I (among other tentative detections). In the present work, we analyze the full data set spanning GRACES' complete wavelength range of 400 to 1050 nm using the Doppler cross-correlation technique \cite[e.g.,][]{Snellen10}. The nominal resolving power of GRACES is $\sim$66,000.

A summary of the observations is presented in Table \ref{tab:obs}. A total of 169 spectra were obtained over the course of the transit (as well as a baseline of observations pre- and post-transit), with approximately 10\% of the transit (lasting 22 minutes) lost when a computer at the observatory crashed and needed to be rebooted. This technical issue did not affect the data preceding or following the gap in the observations. In total, the observations lasted $\sim$5.16 hours. The average signal-to-noise ratio (SNR) ranged from $\sim$22 to $\sim$111 per spectral bin in the orders used in our analysis. The airmass varied between 1.046 and 1.627, and the seeing throughout the observations was excellent, with a measured full-width at half-maximum (FWHM) of 0.55 arcseconds. We refer the reader to \cite{Deibert21b} for additional figures describing the data quality (their Fig.~3).

\begin{deluxetable*}{c c c c c}
\label{tab:obs}
\tablecaption{Summary of GRACES/Gemini North observations of WASP-76b used in this analysis.}
\tablehead{%
    \colhead{Date (UT)} & \colhead{Frames (In/Out)} & \colhead{Exposure Time (s)} & \colhead{{Avg.~SNR (Max.)}\tablenotemark{a}} & 
    \colhead{{Avg.~SNR (Min.)}\tablenotemark{b}}
    }
\startdata
11 October 2020 & 169 (120/49) & 60 & {111.4} & 22.5 \\
\enddata
\tablenotetext{a}{{Average SNR per spectral bin in the 28th order (centered at $\sim$ 807 nm), which had the highest SNR across the full spectrum.}}
\tablenotetext{b}{{Average SNR per spectral bin in the 50th order (centered at $\sim$ 452 nm), which had the lowest SNR of the orders used in this analysis.}}
\end{deluxetable*}

\section{Data Reduction}
\label{sec:reduc}
The initial steps of our data reduction routine proceeded as follows. We extracted the spectra from the raw files using OPERA, the Open source Pipeline for ESPaDOnS Reduction and Analysis \citep{opera}, which performs an optimal extraction, bias subtraction, flat-fielding, blaze correction, continuum normalization, and wavelength calibration. Next, we removed cosmic rays and other outliers using a median absolute deviation flag which masks points greater than 5 median absolute deviations. We then flux-scaled each spectrum \citep[e.g.,][]{Allart17} by dividing out the first spectrum of each night and fitting/dividing out a fourth-order polynomial fit. We note that this is the same process described in \cite{Deibert21b}, albeit across the full GRACES wavelength range in this work; and as in \cite{Deibert21b}, the order of the polynomial fit does not significantly affect our results.

\subsection{Removal of Telluric and Stellar Features with \textsc{SysRem}}
\label{subsec:sysrem}
Following the initial data reduction steps described in Section \ref{sec:reduc}, we corrected for stellar and telluric absorption features using the \textsc{SysRem} algorithm \citep{Tamuz2005}, which is a PCA-like algorithm that removes time-stationary features in a set of spectra. Because the radial velocity of the exoplanet varies significantly throughout the course of its transit (from approximately -52 km/s to +52 km/s in the case of WASP-76b), absorption features from the exoplanet's atmosphere remain intact while essentially time-stationary features from the Earth and the host star are removed by the \textsc{SysRem} algorithm.

Before running \textsc{SysRem}, we interpolated the spectra to a common wavelength grid in the telluric rest frame. We then used the airmass throughout the observations as an initial guess for the first component to be removed by the algorithm. To determine the optimum number of iterations of \textsc{SysRem} to apply to our observations, we ran between 1 and 20 iterations of the algorithm on each order of our spectra. We then repeated our cross-correlation analysis (see Section \ref{subsec:xcorr}) with a high-resolution 1D transmission spectrum generated to match WASP-76b's atmosphere (see Section \ref{subsec:models}) for each number of iterations, and chose the number which maximized the significance of our detection of all species included in the high-resolution transmission spectrum simultaneously (see Fig.~\ref{fig:callie}). 

We found that 7 iterations resulted in the strongest detection, but beyond 2 iterations, the detections were within 1$\sigma$ of each other regardless of the number of iterations. We therefore chose to apply 7 iterations of \textsc{SysRem} to each order of our data. We note that this is one more iteration than used in \cite{Deibert21b}, where we optimized for the Ca~II detection in particular. In the present work, we optimized the algorithm based on a synthetic transmission spectrum which contained absorption features from a number of atoms/molecules across the full GRACES wavelength range. This allowed us to apply the algorithm consistently across every order of the data.

{The results of applying 7 iterations of the \textsc{SysRem} algorithm to each order are presented in Appendix \ref{app:tellurics}.}

\section{Methods}
\label{sec:analysis}
We followed a similar methodology to previous analyses using the Doppler cross-correlation technique \citep[e.g.,][]{Snellen10,Deibert21}. This first involves creating atmospheric models and templates with which to cross-correlate our spectra, and then carrying out the Doppler cross-correlation process on each of these templates.

\subsection{Atmospheric Models and Templates}
\label{subsec:models}
To search for atomic and molecular features in WASP-76b's optical spectrum, we turned to a range of modelling efforts. These are described in further detail below. In particular, we opted for both a custom high-resolution, one-dimensional transmission spectrum created for WASP-76b's atmosphere, as well as a publicly available set of atmospheric templates generated for a generic ultra-hot Jupiter atmosphere. 

In all cases, we prepared the models for cross-correlation with our data by first converting the wavelength grids to air wavelengths, interpolating the wavelength grids to that of GRACES, and convolving the models to the resolution of GRACES using a Gaussian Kernel. We also applied a Butterworth filter to each template in order to mimic the effects of the \textsc{SysRem} algorithm \citep[e.g.,][]{Herman22}.

\subsubsection{High-Resolution 1D Transmission Spectrum}
\label{subsubsec:callie}
We first created a one-dimensional transmission spectrum for WASP-76b's atmosphere based on the parameters described in Table \ref{tab:parameters}. We generated the atmospheric pressure-temperature (\emph{P--T}) profile via the one-dimensional modelling methods described in \cite{Fortney08} and \cite{Fortney20}. Planet-wide average conditions were assumed, as well as equilibrium chemistry at solar metallicity, and the model was iterated to a solution in radiative-convective equilibrium. The \emph{P--T} profile is shown in Fig.~\ref{fig:pt}. We then used the one-dimensional transmission spectrum code described in the appendix of \cite{Morley17} to generate a transmission spectrum at high resolution, based on the calculated \emph{P--T} profile and equilibrium chemical abundances, generally following \cite{Hood20}. We performed these calculations across the wavelength range of GRACES, and included H${}_2$, He, H, H${}_2$O, Fe~I, Na~I, K~I, Li~I, Mn~I, and Ca~II. The resulting spectrum in shown in Fig.~\ref{fig:callie}. {We chose these species in particular as they were expected to be abundant and readily detectable in the atmosphere \citep[e.g.,][]{Kesseli22} had been detected at high significances by multiple previous works \citep[e.g.,][]{Tabernero20,Deibert21b, Kesseli22, SanchezLopez22, AS22}. Due to memory limitations, we could not include additional species in this model; instead, a wider grid of species was explored using the Mantis Network templates \citep[][see below]{mantis}.}

{In order to further compare the results we obtained with our custom model with the generic templates described in the following section, we also generated a one-dimensional transmission spectrum including only Ca~II, H${}_2$, and He. This model is displayed in the left subplot of Fig.~\ref{fig:callie-caii-correlation}.}

\begin{figure}
    \centering
    \includegraphics{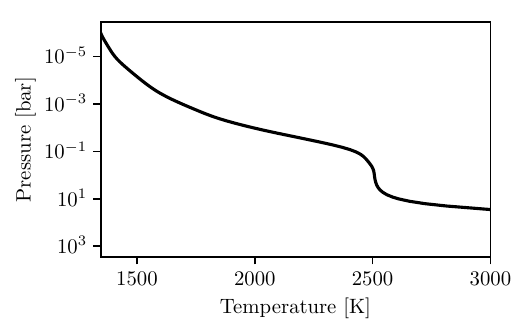}
    \caption{The \emph{P--T} profile generated for WASP-76b, as described in Section \ref{subsubsec:callie}.}
    \label{fig:pt}
\end{figure}

\begin{deluxetable*}{lccc}
\label{tab:parameters}
\tablecaption{Orbital and physical parameters of the WASP-76 system used in this analysis.}
\tablehead{%
    \colhead{Parameter} & \colhead{Symbol (Unit)} & \colhead{Value} & \colhead{Reference}
    }
\startdata
Stellar radius & $R_*$ ($R_\sun$) & $1.756 \pm 0.071$ & E20\tablenotemark{a} \\
Stellar mass & $M_*$ ($M_\sun$) & $1.458 \pm 0.02$ & E20\tablenotemark{ }  \\
Magnitude & $V$ (mag) & $9.52 \pm 0.03$ & \cite{SIMBAD} \\
System scale & $a/R_*$ & $4.08{}^{+0.02}_{-0.06}$ & E20\tablenotemark{ } \\
Orbital period & $P$ (days) & $1.80988198{}^{+0.00000064}_{-0.00000056}$ & E20\tablenotemark{ } \\
Transit duration & $T_{14}$ (min.) & $230$ & E20\tablenotemark{ } \\
Epoch of mid-transit & $T_c$ (BJD) & $2458080.626165{}^{+0.000418}_{-0.000367}$ & E20\tablenotemark{ } \\
Radius ratio & $R_\mathrm{p}/R_*$ & $0.10852 {}^{+0.00096}_{-0.00072}$ & E20\tablenotemark{ }  \\
Planetary radius & $R_\mathrm{p}$ ($R_\mathrm{J}$) & $1.854{}^{+0.077}_{-0.076}$ & \cite{Tabernero20}  \\
Planetary mass & $M_\mathrm{p}$ ($M_\mathrm{J}$) & $0.894{}^{+0.014}_{-0.013}$ & E20\tablenotemark{ } \\
Inclination & $i$ (degrees) & $89.623{}^{+0.005}_{-0.034}$ & E20\tablenotemark{ } \\
Systemic velocity & $\gamma_\mathrm{sys}$ (km/s) & $-1.0733 \pm 0.0002$ & \cite{West16} \\
Stellar radial velocity semi-amplitude & K${}_*$  (m/s) & $116.02{}^{+1.29}_{-1.35}$ & E20\tablenotemark{ } \\
Planetary radial velocity semi-amplitude & K${}_p$ (km/s) & $196.52 \pm 0.94$ & E20\tablenotemark{ } \\
Quadratic limb darkening coefficient & $u_1$ & $0.393$ & E20\tablenotemark{ } \\
Quadratic limb darkening coefficient & $u_2$ & $0.219$ & E20\tablenotemark{ } \\
Projected equatorial rotational velocity & $v\sin i$ (km/s) & 1.48 $\pm$ 0.28 & E20\tablenotemark{ } \\
\enddata
\tablenotetext{a}{\cite{Ehrenreich20}}
\end{deluxetable*}

\begin{figure*}
\centering
\includegraphics[width=\textwidth]{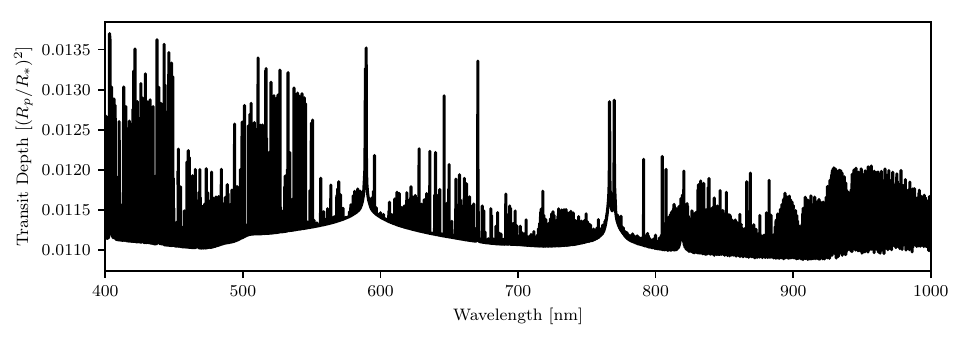}
\caption{A high-resolution, one-dimensional synthetic transmission spectrum generated for WASP-76b's atmosphere using the methods described in Section \ref{subsubsec:callie}. A number of spectral features can be seen, including a band of strong Fe~I features at the blue end of the spectrum, strong absorption features due to Na~I at $\sim$589 nm, and many absorption lines due to H${}_2$O at the red end of the spectrum.}
\label{fig:callie}
\end{figure*}

\subsubsection{Mantis Network Templates}
In addition to the model described above, we also searched for atmospheric absorption using a standard grid of templates from the Mantis Network \citep{mantis}. As described in \cite{mantis}, the grid includes high-resolution templates for more than 140 atmospheric species across a range of atmospheric temperatures. While the public database currently only includes templates generated for ultra-hot (i.e., $>$ 2000~K) atmospheres observed in transmission, the full database will eventually include templates for lower temperatures as well \citep{mantis}. {The goal of the Mantis Network database is to offer a standardized set of atmospheric templates which can be used with a range of different spectrographs in order to make analyses using different instruments more consistent. We therefore chose to use these templates as a way to better compare our work to recent analyses of WASP-76b using data from ESPRESSO.}

The atmospheric templates from the Mantis Network were recently used by \cite{AS22} to analyze transmission spectra of WASP-76b from ESPRESSO, resulting in a new detection of Ba~II, among other previously detected species \citep{AS22}. {\cite{AS22} also used these same templates to detect a number of species in the atmosphere of WASP-121b, while \cite{Borsato23} used the Mantis templates with archival observations of KELT-9b from HARPS-N and CARMENES.} 
{With the present work, our aim is to both compare our analysis with that of \cite{AS22} while also demonstrating the efficacy of the Mantis templates in detecting atmospheric species with GRACES spectra.}

We refer the reader to \cite{mantis} for a {full} description of the template generation. {Briefly, the templates are created assuming a generic ultra-hot Jupiter with a surface gravity of $g = 2000$ cm s${}^{-1}$, a planetary radius of $R_p = 1.5$ $R_J$ at a pressure of 10 bar, and an isothermal pressure-temperature profile throughout the atmosphere at temperatures of 2000~K, 2500~K, 3000~K, 4000~K or 5000~K. As described in \cite{mantis}, the fact that the cross-correlation template is normalized means that it is largely insensitive to the choices of $g$ and $R_p$, assuming that the atmospheric scale height $H$ is much smaller than $R_p$. Furthermore, \cite{mantis} demonstrated that varying the surface gravity $g$ does not affect the final detection SNR at greater than the 0.5\%-level.}

{The templates used in this work included continuum-forming species and a single line-forming species. The continuum opacity sources considered in the calculation of these templates included collision-induced absorption (CIA) of H${}_2$-H${}_2$, H${}_2$-He, and H-He collisions; free-free and bound-free absorption of H${}^-$ and H${}^{-}_{2}$; and Rayleigh scattering cross-sections for H${}_2$, H, and He \citep{mantis}. The templates were calculated assuming equilibrium chemistry and solar elemental abundances.}

For the present work, we downloaded the templates generated for 2000~K, 2500~K, 3000~K, and 4000~K atmospheres. While the 2000~K atmosphere is closest to WASP-76b's expected equilibrium temperature of $\sim$2200~K \citep[][]{West16}, \cite{Deibert21b} demonstrated that the atmospheric layers probed by these GRACES observations may well be much hotter than the equilibrium temperature. {Furthermore, a number of recent studies of WASP-76b have found temperatures ranging from $\sim$2000 K to $\sim$4000 K. For example, \cite{Landman21} retrieve a temperature of between 2700 K and 3700 K, while \cite{Seidel21} retrieve a temperature of 3389 K (though they note that this is unusually high). Yet analyses of HST data have yielded temperatures of 2300 K \citep{vonEssen20} and 2231 K \citep{Edwards20}. \cite{Kesseli22} and \cite{CB21} also found that temperatures varying between $\sim$2000 K and $\sim$4000 K yielded similar final results in their analyses.}
We thus carried out our analysis for a range of atmospheric temperatures. Note that in their analysis of WASP-76b, \cite{AS22} made use of the 2500~K atmospheric templates.

We ran a search to determine which of the templates contained absorption lines within the wavelength ranges of each GRACES order. {We did this by dividing the templates into wavelength ranges corresponding to each GRACES order, and eliminating wavelength ranges for which the corresponding transit depth was zero everywhere. We fully eliminated models which had zero transit depth throughout the full GRACES wavelength range. This yielded a total of 81 templates.}
While we carried out our analysis for all these species, for the sake of brevity we only present detailed discussions for species which (i) yielded a detection in WASP-76b's atmosphere in the present work, or (ii) had previously been detected in WASP-76b's atmosphere.

\subsection{Doppler Cross-Correlation}
\label{subsec:xcorr}
To search for atmospheric absorption features in our data, we made use of the Doppler cross-correlation technique \cite[e.g.,][among many others]{Deibert21,Snellen10}. Briefly, this method involves Doppler-shifting atmospheric models or templates to a range of radial velocities (RVs), and then cross-correlating the shifted models with each spectrum, in order to extract the planetary signal from the data as the planet transits the host star. While a very strong signal can be seen by eye in this correlation map, it's often necessary to boost the strength of this signal by then phase-folding to a range of Keplerian velocities, $K_p$. The planetary signal is then visible as a correlation peak in the 2D $K_p$-RV map.

We ran an automated search on each template used in our analysis to determine whether the template contained spectral features in each GRACES order. Orders which did not include any spectral features for a given template were excluded from the cross-correlation analysis for that template. As an example of this, see Fig.~\ref{fig:CaII}, which displays a Ca~II atmospheric templates used in one of our cross-correlations. The template only contains absorption features in a small region of the full GRACES wavelength range, and orders which did not contain absorption features were therefore not included in that template's cross-correlation analysis.

To estimate the detection significance, we followed e.g., \cite{Boucher21} in dividing out the standard deviation of the $K_p$-RV map calculated by excluding a region of $\pm$30 km/s in both $K_p$ and RV around the peak correlation. \cite{Boucher21} demonstrated that this is a robust, computationally efficient method of determining the significance of a cross-correlation detection. {Note that we assume that the distribution of the cross-correlation function is Gaussian \citep[e.g.,][]{Brogi12,Birkby18}.}

{We chose 30 km/s as a threshold for calculating the standard deviation because a visual inspection of the results indicated that this would sufficiently cover even the broadest detected signals. However, to confirm that the chosen threshold does not impact our final results, we tested the effects of varying this threshold from 1 km/s to 50 km/s and recalculating the final SNR. We found that in almost all cases, the final SNR does not vary by more than 1$\sigma$ regardless of the threshold used. The exception is Ca~II, which varied by $\sim$1.5$\sigma$ when the threshold was increased. Due to the very broad nature of the Ca~II signal, however, much of this variation came from smaller thresholds which include a large portion of the cross-correlation peak in the standard deviation calculation. The variation in the final SNR for thresholds which fully encompass the Ca~II peak was $<$ 1$\sigma$, as was the case for the other detected signals. We therefore conclude that a threshold of 30 km/s is sufficient and does not affect our final results.}

\section{Results and Discussion}
\label{sec:results}
\subsection{Detections}
\label{subsec:detections}

\subsubsection{Detection of WASP-76b's Atmosphere with the 1D Transmission Spectra}
We began by carrying out the Doppler cross-correlation technique with the one-dimensional transmission spectrum created for WASP-76b's atmosphere. Given that the spectrum contained a number of species which had previously been detected in WASP-76b's atmosphere \citep[e.g., Fe~I and Ca~II;][]{Ehrenreich20,Deibert21b}, we expected the synthetic transmission spectrum to correlate strongly with our data. The model contains absorption features across the full wavelength range of GRACES; however, we excluded orders with particularly low quality data from our analysis. {We determined these low quality regions by excluding orders for which the average SNR per order across all spectra was below 20. At these low SNRs, the spectra extracted by the OPERA pipeline contained unphysical negative values, which in turn resulted in \texttt{nan} values in the final \textsc{SysRem}-reduced spectra. In total, this threshold resulted in the exclusion of 6 orders at the blue end of the spectrum for which the average SNR was below 20.}

\begin{figure*}
\centering
\includegraphics[width=\textwidth]{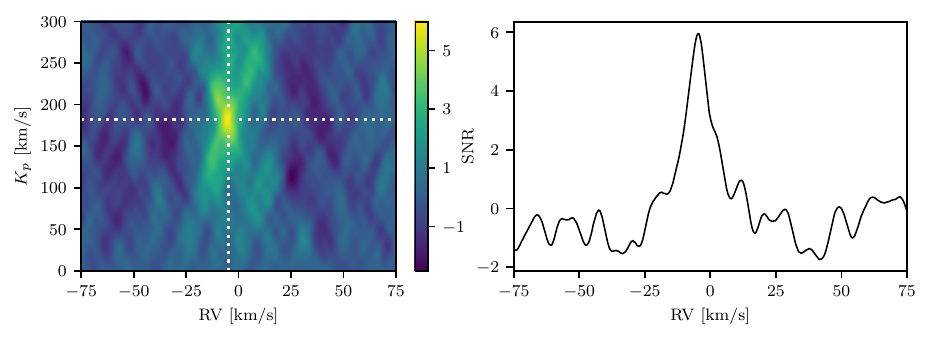}
\caption{The results of cross-correlating the one-dimensional synthetic transmission spectrum generated for WASP-76b's atmosphere with our observations. The plot on the left shows the 2D $K_p$-RV map, with the peak correlation identified by the dotted white lines, and the SNR indicated by the colorbar. The plot on the right shows a slice of the 2D map at the $K_p$ value corresponding to the peak correlation. The atmosphere is detected at a significance of 6.0$\sigma$, with a measured $K_p$ of $182^{+12}_{-15}$ km/s and a measured $V_{\text{center}}$ of $-4.6 \pm 2.1$ km/s. The latter could be interpreted as atmospheric winds.}
\label{fig:callie-correlation}
\end{figure*}

The results of this cross-correlation are presented in Fig.~\ref{fig:callie-correlation}. When correlating our data with this model, we are able to detect WASP-76b's atmosphere at a significance of 6.0$\sigma$. The peak SNR of the cross-correlation function is located at a Keplerian velocity of $K_p = 182^{+12}_{-15}$ km/s, where we have taken as error the point at which the peak SNR in the cross-correlation function drops by 1$\sigma$. This is lower than (though within error of) the value reported by \cite{Ehrenreich20}, but consistent with those determined in \cite{Deibert21b}, using the same data but a different methodology. The peak SNR is also located at a $V_{\text{center}}$ value of $-4.6\pm 2.1$ km/s. This value could be interpreted as the exoplanet's atmospheric wind speed, and is consistent with those derived for various atmospheric species in \cite{Deibert21b}, although we note that the errors are large and we therefore caution against interpreting this as a true representation of the planet's atmospheric wind. As before, we have taken as error the point at which the peak SNR drops by 1$\sigma$.

The species used to create the synthetic transmission spectrum have previously been detected in the atmosphere of WASP-76b \citep[H${}_2$O, Fe~I, Na~I, K~I, Li~I, Mn~I, Ca~II;][]{SanchezLopez22,Ehrenreich20,Seidel19,Tabernero20,Kesseli22,Deibert21b}, though we note that H${}_2$O has only been detected at near-infrared wavelengths. It is thus likely that all of these species are contributing to the cross-correlation signal. 

\begin{figure*}
    \centering
    \includegraphics{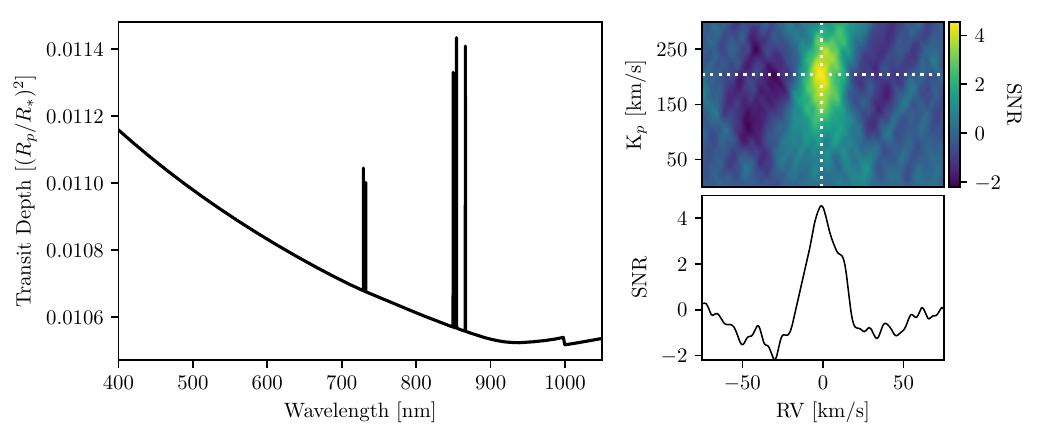}
    \caption{The results of cross-correlating the one-dimensional synthetic transmission spectrum containing only Ca~II, H${}_2$, and He generated for WASP-76b's atmosphere with our observations. The plot on the left shows the high-resolution, one-dimensional synthetic transmission spectrum generated using the methods described in Section \ref{subsubsec:callie}. The plot on the top right shows the 2D $K_p$-RV map, with the peak correlation identified by the dotted white lines and the SNR indicated by the colorbar. The plot on the bottom right shows a slice of the 2D map at the peak $K_p$ value. We detect Ca~II in the atmosphere with this model at a significance of 4.5$\sigma$, with a measured $K_p$ of 204${}^{+35}_{-46}$ km/s and a measured $V_\mathrm{center}$ of ${-1.0}_{-4.8}^{+4.5}$ km/s.}
    \label{fig:callie-caii-correlation}
\end{figure*}

{In order to better compare our custom models with results from the more generic Mantis templates (see the following section), we also created a one-dimensional transmission spectrum containing only Ca~II, H${}_2$, and He, and repeated the cross-correlation process described above. The results of this are presented in Fig.~\ref{fig:callie-caii-correlation}. We detect Ca~II in the atmosphere with this model at a significance of 4.5$\sigma$, with a measured $K_p$ of 204${}^{+35}_{-46}$ km/s and a measured $V_\mathrm{center}$ of ${-1.0}_{-4.8}^{+4.5}$ km/s. These results are discussed and compared to the Mantis results in more detail in Section \ref{subsec:callievmantis}.}

In the following section, we look at individual species separately through cross-correlations with atmospheric templates from the Mantis Network \citep{mantis}.

\subsubsection{Detections of Individual Species with the Mantis Network Templates}
\begin{figure*}[t]
    \centering
    \includegraphics[]{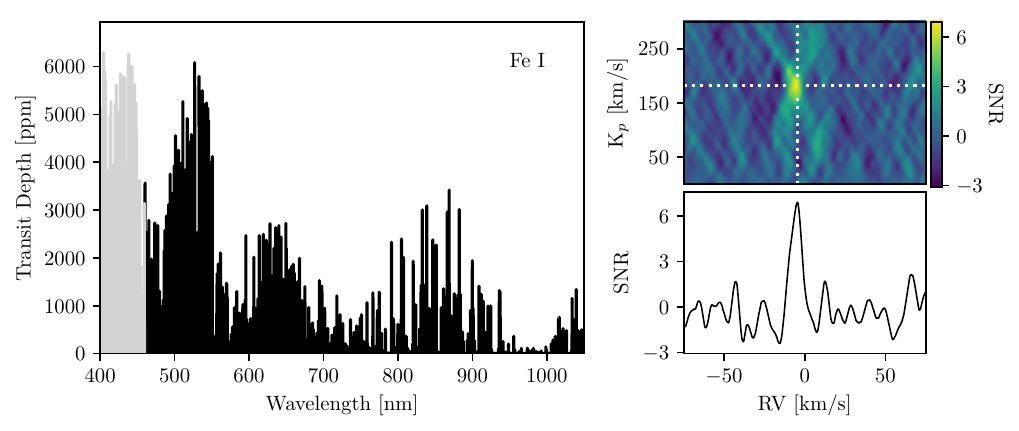}
    \caption{A detection of Fe~I absorption in WASP-76b's atmosphere. \textit{Left}: the Mantis Network template cross-correlated with our spectra. The region plotted in grey was excluded due to the low quality of the data at the edge of GRACES' wavelength range. The 2500~K template is used. \textit{Top Right}: The 2D cross-correlation map. The white lines indicate the peak SNR, which corresponds to a $>5$$\sigma$ detection of Fe~I in WASP-76b's atmosphere. \textit{Bottom Right}: A slice of the 2D cross-correlation map at the $K_p$ corresponding to the peak SNR. The signal is slightly offset from RV = 0 km/s, indicative of winds in the exoplanet's atmosphere.}
    \label{fig:FeI}
\end{figure*}

\begin{figure*}[t]
    \centering
    \includegraphics[]{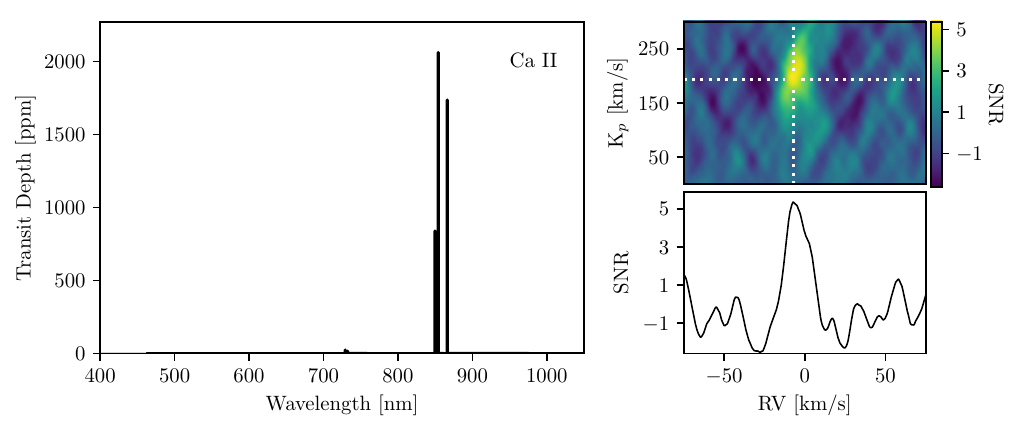}
    \caption{A detection of Ca~II absorption in WASP-76b's atmosphere. The plots are as described in the caption of Fig.~\ref{fig:FeI}. The 2000~K template is used. Only orders containing Ca~II lines were used in the analysis, meaning that no orders were excluded due to low data quality.}
    \label{fig:CaII}
\end{figure*}

\begin{figure*}
    \centering
    \includegraphics[]{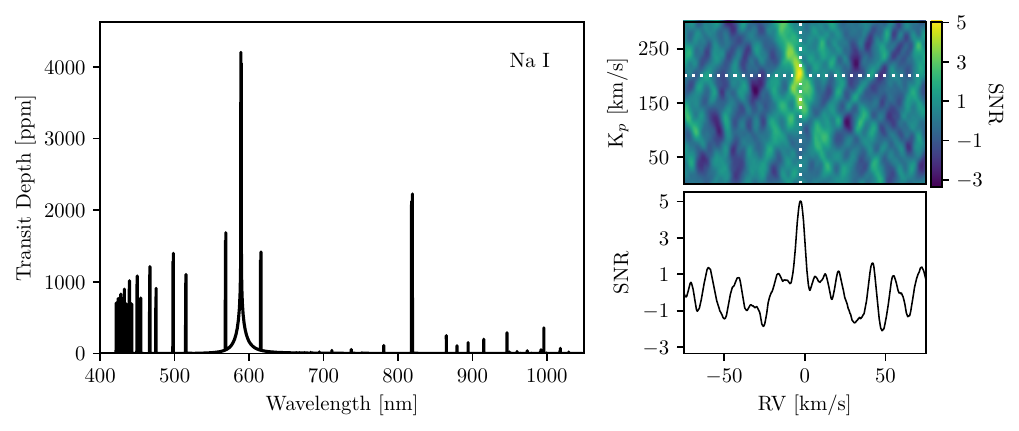}
    \caption{A detection of Na~I absorption in WASP-76b's atmosphere. The plots are as described in the caption of Fig.~\ref{fig:FeI}. The 2000~K template is used. Due to the small number of lines present in the model, we opted to also include lines from the blue end of the spectrum. When these lines are excluded, the detection significance falls below 5$\sigma$.}
    \label{fig:NaI}
\end{figure*}

Using the Mantis Network templates \citep{mantis}, we detect Fe~I, Ca~II, and Na~I at high significance ($>$5$\sigma$) in the atmosphere of WASP-76b. These detections, along with the Mantis templates cross-correlated with our spectra, are presented in Figs.~\ref{fig:FeI}, \ref{fig:CaII}, and \ref{fig:NaI}. As discussed earlier, these species have all previously been detected in the atmosphere of WASP-76b. We briefly discuss each of these detections below. {Note that while the template with the temperature corresponding to the strongest detection is presented in each figure, we provide all detection significances (which in most cases are within $<$1$\sigma$ of each other) in Appendix \ref{app:sigs}. We caution that while we have chosen to display different temperature templates in these figures, our analysis does not include a true retrieval of the atmospheric temperature and should not be interpreted as such.}

Fe~I absorption was first reported by \cite{Ehrenreich20}, and later confirmed by \cite{Tabernero20}, \cite{Kesseli21}, \cite{Kesseli22}, and \cite{AS22}. In the present work, we detect Fe~I absorption via cross-correlation with the 2500~K Mantis Network template at a significance of 6.9$\sigma$. When the 2000~K template is used, this significance falls to 6.6$\sigma$, while the 3000~K template yields a significance of 6.8$\sigma$ and the 4000~K template yields a significance of 6.0$\sigma$. While this could indicate that the atmospheric layers probed by Fe~I are hotter than the equilibrium temperature, and fall somewhere between 2500 and 3000~K, we caution against reading too much into this result. More sophisticated modelling will be necessary in order to accurately account for all factors which may impact the temperature \citep[for e.g., NLTE effects;][]{Fossati20,Fossati21,Deibert21b} and place constraints on the temperature of the layers probed.

In their initial analysis of the Fe~I signal in WASP-76b's atmosphere, \cite{Ehrenreich20} reported asymmetry in the signal between the first and second halves of the transit (a result which has been interpreted as condensation and ``iron rain'', though see \cite{Wardenier21} and \cite{Savel22} for further interpretations). Yet due to the lower SNR and resolution of our data, as well as the fact that 10\% of the second half of the transit was lost due to technical issues at the observatory, we are unable to draw substantive conclusions about this asymmetric signal in the present work.

Using this same 2500~K Fe~I template, we find a Keplerian velocity of $K_p = 182^{+8}_{-14}$ km/s, in line with our previous results \citep[][]{Deibert21b} and consistent within error with the value reported by \cite{Ehrenreich20}. We also find a wind speed of $V{}_{\text{center}} = -4.4^{+0.9}_{-2.1}$ km/s which, again, is consistent with our previous results \citep[][]{Deibert21b} and matches well with the value we find for the one-dimensional synthetic transmission spectrum above.

We note that we also searched for Fe~II absorption (as the exoplanet's high atmospheric temperatures could ionize Fe~I) but did not detect it with any of the templates. This is discussed in further detail in Section \ref{subsec:non}.

Ca~II was previously detected in WASP-76b's atmosphere by \cite{CB21} and \cite{Deibert21b}. While we resolved the individual lines of the Ca~II triplet via transmission spectroscopy in \cite{Deibert21b}, we have now shown that Ca~II is also readily detectable via Doppler cross-correlation with GRACES spectra, even when only a few lines are present in the atmospheric template (see the left side of Fig.~\ref{fig:CaII}). 

Using this technique, we detect atmospheric Ca~II at a significance of 5.3$\sigma$ when correlated with the 2000~K template. The significance drops to 4$\sigma$ when we use the 2500~K template, and atmospheric Ca~II is not detectable with the 3000~K template. Interestingly, the significance rises again with the 4000~K template, to 4.2$\sigma$. Fig.~\ref{fig:CaII}  displays the 2000~K template and results. While this could indicate that the temperature of the atmospheric layer probed by the Ca~II lines falls somewhere between 2000 and 2500~K, we again caution against reading too much into this result. In particular, we note that \cite{CB21} found the greatest correlation strength with CARMENES spectra for a 4000~K Ca~II model, rather than a 2200 K model matching the planet's equilibrium temperature. However, their model was created to match the specific parameters of the WASP-76b system (unlike the generic Mantis templates used in this work), and also included the effects of tidally-locked rotation. Even so, the differences in correlation strengths between their four models (2200 K and 4000 K with and without tidally-locked rotation) were not significant, and varied between $\sim$6$\sigma$ and $\sim$8$\sigma$. We conclude that more sophisticated modelling will be necessary in order to truly constrain the atmospheric temperature.

Using this 2000~K template, we find a Keplerian velocity of $K_p = 193^{+27}_{-17}$ km/s, which is consistent with what we found in \cite{Deibert21b}, though we note that the errors in this case are particularly large due to how broad the signal is in the $K_p$-RV map. We also recover a wind speed of $V_{\text{center}} = -7.1^{+5.1}_{-2.7}$ km/s. We note that this is a greater blueshift than reported in \cite{Deibert21b}; however, the results are consistent within error (though again, the errors in this case are particularly large). The large errors in our recovered values are likely due to the fact that very few lines were used in this cross-correlation. Nevertheless, we have shown that GRACES is adept at detecting atmospheric Ca~II through cross-correlation, if not at providing stringent constraints on the Keplerian velocity and atmospheric winds. This could be due to the fact that we searched for the Ca~II lines directly in \cite{Deibert21b}, whereas in the present work we cross-correlated with a generic template which may not match the true line widths/depths of the Ca~II triplet. {Additionally, because we are effectively averaging the three lines of the Ca~II triplet with the cross-correlation method and searching for all three lines simultaneously, errors from the noisiest line (at $\sim$850 nm, located close to the edge of its spectral order) may increase the overall errors in our detection as compared to the method presented in \cite{Deibert21b}, which resolves the three lines individually.}

Na~I was the first species to be detected in WASP-76b's atmosphere \citep{Zak19, Seidel19}, and has been recovered in a number of later studies \citep{Tabernero20,Seidel21,Deibert21b,Kesseli22,AS22}. In this work, we detect Na~I at a significance of 5.0$\sigma$ when using a 2000~K template. With the 2500~K template, this drops to 4.2$\sigma$. Interestingly, the significance rises slightly with the 3000~K template, yielding a 4.7$\sigma$ detection; the 4000~K template yields a 4.1$\sigma$ detection. We display the 2000~K template in Fig.~\ref{fig:NaI}. With this template, we find $K_p = 201^{+11}_{-12}$ km/s. Again, the errors are particularly large; this may be due to the small number of lines used in the cross-correlation. We also determine a wind speed of $V_{\text{center}} = -2.6^{+1.5}_{-1.8}$ km/s. This is consistent within error with the results from \cite{Deibert21b}.

The results from this section are summarized in Table \ref{tab:detections-5}.

\begin{deluxetable*}{ccccc}
\label{tab:detections-5}
\tablecaption{Summary of detected ($>$5$\sigma$) and tentatively detected (5$>$$\sigma$$>$3) species.}
\tablehead{%
    \colhead{Species} & \colhead{Significance ($\sigma$)} & \colhead{Temperature (K)} & \colhead{$K_p$ (km/s)} & \colhead{$V_{\text{center}}$ (km/s)}
    }
\startdata
Fe~I & 6.9 & 2500 & $182^{+8}_{-14}$ & $-4.4^{+0.9}_{-2.1}$ \\
Ca~II & 5.3 & 2000 & $193^{+27}_{-17}$ & $-7.1^{+5.1}_{-2.7}$ \\
Na~I & 5.0 & 2000 & $201^{+11}_{-12}$ & $-2.6^{+1.5}_{-1.8}$ \\
\hline
Li~I & 4.2 & 2500 & $221^{+17}_{-14}$ & $0.7 \pm 2.4$ \\
K~I & 4.4 & 3000 & $209^{+29}_{-6}$ & $-5.6 \pm 1.2$ \\
Cr~I & 4.0 & 3000 & $208^{+8}_{-11}$ & $-3.2^{+1.8}_{-1.2}$ \\
V~I & 4.6 & 4000 & $140^{+13}_{-16}$ & $-7.4^{+1.8}_{-2.1}$ \\
\enddata
\tablecomments{Column 1: species detected. Column 2: significance of detection. Column 3: temperature of atmospheric template which yields the highest significance. \textbf{N.B.} we do not claim that this is indeed the atmospheric temperature at the layers probed. Column 4: Keplerian velocity. Column 5: offset from expected line location, assuming a Gaussian profile. In some cases, this could be attributed to atmospheric winds.}
\end{deluxetable*}

\subsection{Tentative Detections}
\label{subsec:tentative}
In addition to the results presented above, a number of Mantis Network templates yielded detections with significances below 5$\sigma$ yet above 3$\sigma$. We classify these results as tentative detections warranting follow-up studies (although we note that all of these species have previously been detected in the atmosphere of WASP-76b, meaning that in this case a follow-up study with GRACES is not necessary in order to confirm their presence). {Note again that while the template corresponding to the strongest tentative detection is presented in each figure, we provide all detection significances in Appendix \ref{app:sigs}.}

\begin{figure*}
    \centering
    \includegraphics[]{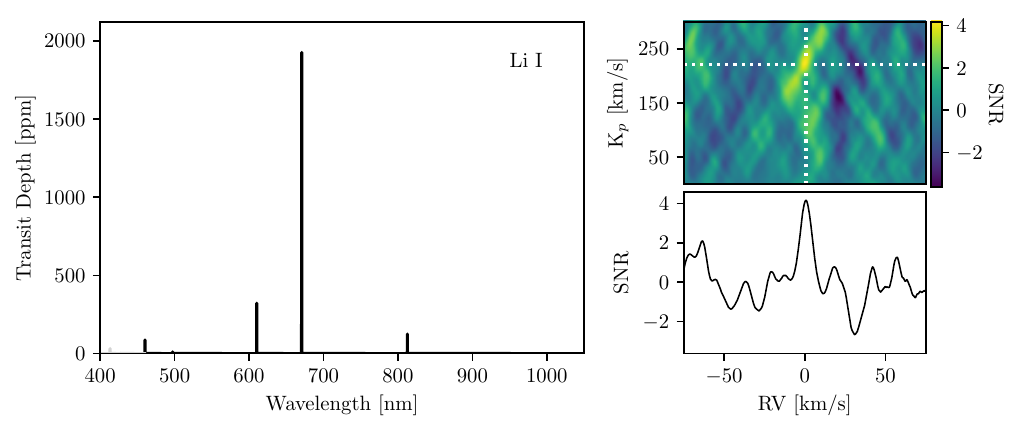}
    \caption{A tentative detection of Li~I absorption in WASP-76b's atmosphere. The plots are as described in the caption of Fig.~\ref{fig:FeI}. While we do see a peak at 4.2$\sigma$ near the expected location of the exoplanet's atmosphere, the noise in the cross-correlation map prevents us from confidently labelling this as a detection. The 2500~K template was used, and 6 orders with low data quality containing only weak Li~I lines at the blue end of the spectrum were excluded (shown in grey).}
    \label{fig:LiI}
\end{figure*}

\begin{figure*}
    \centering
    \includegraphics[]{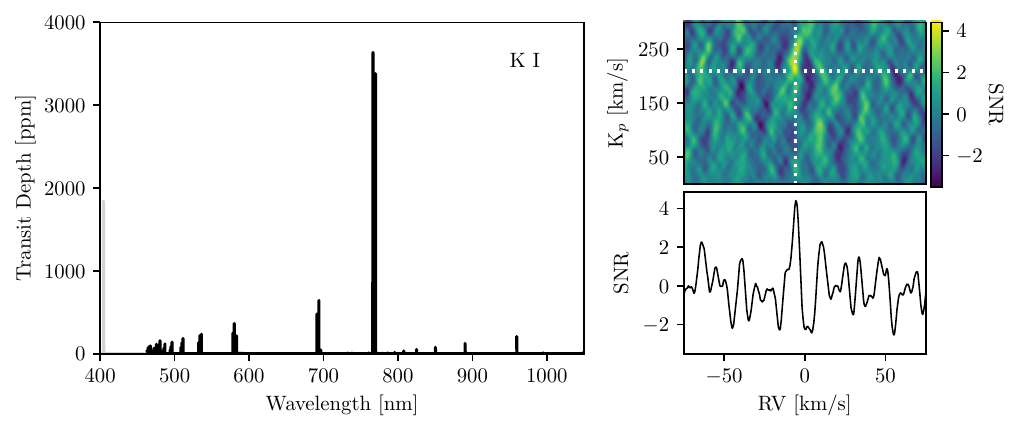}
    \caption{A tentative detection of K~I absorption in WASP-76b's atmosphere. The plots are as described in the caption of Fig.~\ref{fig:FeI}. While we do see a peak at 4.4$\sigma$ at roughly the expected location of the exoplanet's atmosphere, the noise in the cross-correlation map prevents us from confidently labelling this as a detection. The 3000~K template was used. The K~I line at the blue end of the spectrum was excluded due to the low quality of the data (shown in grey).}
    \label{fig:KI}
\end{figure*}

\begin{figure*}
    \centering
    \includegraphics[]{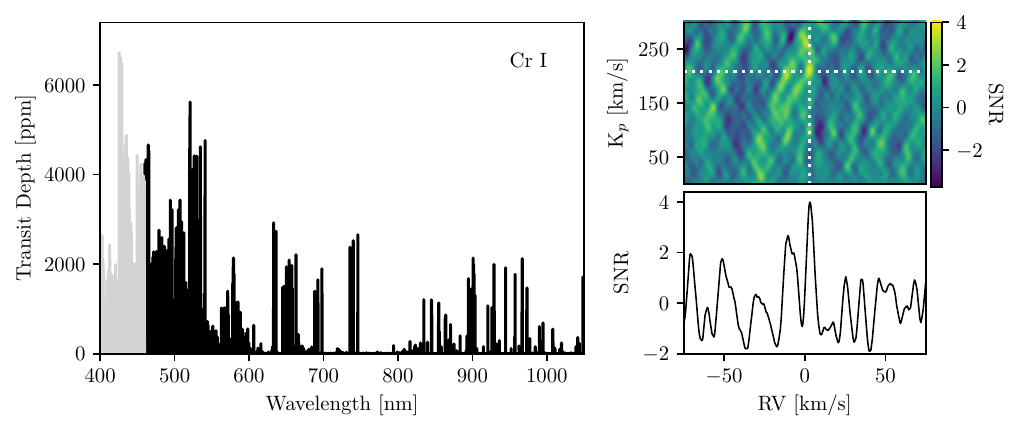}
    \caption{A tentative detection of Cr~I absorption in WASP-76b's atmosphere. The plots are as described in the caption of Fig.~\ref{fig:FeI}. While we do see a peak at 4.0$\sigma$ at roughly the expected location of the exoplanet's atmosphere, the noise in the cross-correlation map prevents us from confidently labelling this as a detection. The 3000~K template was used. Six orders at the blue end of the spectrum were excluded from the analysis (indicated in grey), similar to the 1D high-resolution synthetic transmission spectrum and the Fe~I model.}
    \label{fig:CrI}
\end{figure*}

\begin{figure*}
    \centering
    \includegraphics[]{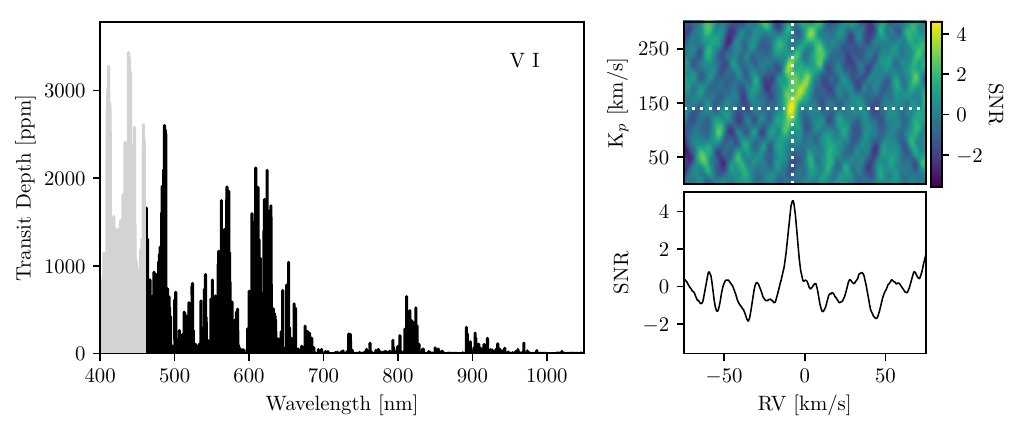}
    \caption{A tentative detection of V~I absorption in WASP-76b's atmosphere. The plots are as described in the caption of Fig.~\ref{fig:FeI}. The 4000~K template was used. Six orders at the blue end of the spectrum were excluded from the analysis (indicated in grey), similar to the 1D high-resolution synthetic transmission spectrum and the Fe~I model.}
    \label{fig:VI}
\end{figure*}

The species tentatively detected in WASP-76b's atmosphere include Li~I (Fig.~\ref{fig:LiI}), K~I (Fig.~\ref{fig:KI}), Cr~I (Fig.~\ref{fig:CrI}), and V~I (Fig.~\ref{fig:VI}). The former two species were also tentatively detected in \cite{Deibert21b} using this same dataset, albeit different methodology, and tentatively detected by \cite{CB21} using CARMENES/Calar Alto. Li~I was detected by \cite{Tabernero20}, \cite{Kesseli22}, and \cite{AS22} using different instruments, while K~I was detected by \cite{Tabernero20} but only tentatively detected by \cite{Kesseli22}. On the other hand, V~I and Cr~I were not considered in \cite{Deibert21b}, but were previously detected by \cite{Kesseli22} and \cite{AS22}.

In the case of Li~I, the 2500~K template yields the highest significance, corresponding to a 4.2$\sigma$ tentative detection. However, we note that there are a number of additional peaks present at high significances. The Keplerian velocity is higher than those recovered for other species, with $K_p = 221^{+17}_{-14}$ km/s, while the wind speed is $V_{\text{center}} = 0.7 \pm 2.4$ km/s. 

In the case of K~I, we find the highest significance for the 3000~K template, which is detected at 4.4$\sigma$. However, as can be seen in Fig.~\ref{fig:KI}, there are a number of peaks approaching 3$\sigma$ throughout the $K_p$-RV map, which calls this result into question. Nevertheless, we are able to extract a Keplerian velocity of $K_p = 209^{+29}_{-6}$ km/s and a wind speed of $V_{\text{center}} = -5.6 \pm 1.2$ km/s. While K~I has already been detected or tentatively detected by several other studies \citep{Tabernero20,CB21, Deibert21,Kesseli22}, the results presented here have likely been affected by an imperfect telluric correction (note that the strongest K~I lines fall within a forest of dense O${}_2$ absorption; {see also Appendix \ref{app:tellurics} for a visualization of the efficacy of our telluric removal}) and should be treated with particular scrutiny. We note as well that both the 2500~K and 4000~K templates yield very similar significances of 4.3$\sigma$.

For Cr~I, we find the highest significance using the 3000~K model, which we detect at 4.0$\sigma$. Yet similarly to K~I, there are many additional peaks at the 3$\sigma$ level in the $K_p$-RV map, meaning that this result should also be treated with scrutiny. The peak significance is located at a Keplerian velocity of $K_p = 208^{+8}_{-11}$ km/s and a positive (i.e., redshifted) wind speed of $V_{\text{center}} = 3.2^{+1.8}_{-1.2}$ km/s. This is different from the other $V_{\text{center}}$ results presented in this work, and further indicates that this tentative detection should be considered with particular scrutiny.

Finally, we detect V~I at a significance of 4.6$\sigma$ using the 4000~K template. The Keplerian velocity is much lower than that detected for other species, with $K_p = 140^{+13}_{-16}$ km/s, while the atmospheric wind speed is $V_{\text{center}} = -7.4^{+1.8}_{-2.1}$ km/s. Again, we caution that this result is tentative. We also note that this is the only detected (or tentatively detected) species which has the strongest correlation with the 4000~K template. This could indicate that our V~I observations are probing a hotter region of the atmosphere{; however, the differences between the calculated significances for the four templates are small (4.2$\sigma$ for the 3000~K template, 4.0$\sigma$ for the 2500~K template, and 3.6$\sigma$ for the 2000~K template) and likely not significant enough to determine the temperature concretely.}

The tentative detections described in this section are outlined in Table \ref{tab:detections-5}. {As can be seen in the figures presented in this section, there are a number of spurious peaks at $K_p$ and $V_\mathrm{center}$ values away from the peak correlation which are largely limiting our ability to classify these as definite detections. As demonstrated in \cite{Esteves17}, spurious peaks $>$1$\sigma$ can arise in cross-correlations, particularly in cases where the templates contain many closely spaced lines. Combining our observations with additional observations from ESPRESSO or HARPS would help reduce the contribution of these spurious peaks; however, such an analysis is beyond the scope of this work.}

\subsection{Non-Detections and Model Injection/Recovery Tests}
\label{subsec:non}
The remaining Mantis Network templates analyzed in this work did not result in significant ($>5\sigma$) or even tentative ($3 < \sigma < 5$) detections. For the sake of brevity, we have only included detailed plots for a subset of these species. The 2D cross-correlation maps for the rest are presented in Appendix \ref{app:nondetec-5}.

In some cases, there were species detected (or tentatively detected) by previous studies that we were not able to recover with our analysis. These include Mg~I, Mn~I, Ni~I, Sr~II, Ba~II, and Co~I. Mn~I and Mg~I were detected by \cite{Tabernero20}, \cite{Kesseli22}, and \cite{AS22}; Ni~I and Sr~II were detected by \cite{Kesseli22}; Ba~II was detected by \cite{AS22}; and Co~I was tentatively detected by \cite{Kesseli22}. Notably, the detections in \cite{AS22} were made using the same Mantis Network templates that we employed in the present work \citep{mantis}.

To investigate these non-detections further, we carried out model injection/recovery tests for each of the above species. We also included Ti~I, Ti~II, Ca~I, Al~I, Sc~I, Sc~II, and Zr~I, which \cite{Kesseli22} note should be readily observable in the atmosphere of WASP-76b (at the wavelength range covered by ESPRESSO) but which we did not detect in our analysis; H${}_2$O, which has been detected in the near-infrared in the atmosphere of WASP-76b \citep{SanchezLopez22} but which we did not detect in the optical; and Fe~II, to investigate whether or not ionized Fe~I would be detectable in our observations given that Fe~I was detected.

To carry out these tests, we injected each atmospheric template into our data with a Keplerian velocity equivalent to the negative of the planet's Keplerian velocity. We injected the templates at this negative value, rather than at the planet's true velocity, in order to avoid boosting any weak absorption that may be present at the planet's true velocity. We then repeated the Doppler cross-correlation process described in Section \ref{subsec:xcorr}, albeit searching for signals at $-K_p$. For each species, we carried out this injection/recovery process using the 2000~K, 2500~K, {3000~K, and 4000~K} templates. We note that because we are cross-correlating with the same templates that have been injected into the data, and it is unlikely that the true atmosphere is identical to these templates, our model injection/recovery tests are likely overestimating our detection sensitivities.

In the cases of Sr~II and Al~I, we were unable to recover any of the injected models. This suggests that even if these species were present in the atmosphere of WASP-76b, our GRACES spectra would not be sensitive enough to detect them. Notably, this explains why we were unable to detect Sr~II, which was previously detected in the atmosphere of WASP-76b \citep{Kesseli22}. 

\begin{figure*}
    \centering
    \includegraphics[]{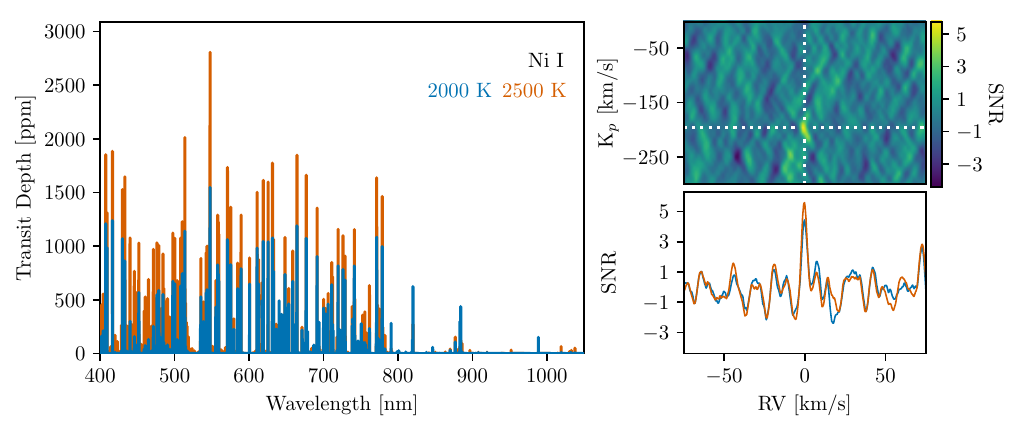}
    \caption{The model injection/recovery process for Ni~I. \textit{Left:} the Mantis Network templates injected and cross-correlated with our data. The 2000~K template is shown in blue, and the 2500~K in orange. The 3000~K and 4000~K templates are omitted because there are no lines present at the GRACES wavelength range. \textit{Top Right:} The 2D cross-correlation map for the template recovered at the highest significance, which in this case is the 2500~K template. The white lines indicate the peak SNR, which is also the velocity at which the template was injected into our data. \textit{Bottom Right:} Slices of the 2D cross-correlation maps for each of the templates, with the 2000~K template shown in blue and the 2500~K template shown in orange as in the plot on the left. The 2500~K template is recovered at a significance of $>$5$\sigma$, while the 2000~K template is only tentatively recovered.}
    \label{fig:NiI}
\end{figure*}

In the case of Ni~I, which was previously detected by \cite{Kesseli22}, we are able to recover the 2500~K template at 5.7$\sigma$ but unable to significantly detect the injected 2000~K template (although we do tentatively recover it, at a significance of 4.5$\sigma$). We note that there are no lines present in the 3000~K or 4000~K Mantis Network templates at the GRACES wavelength range. In their analysis, \cite{Kesseli22} used an isothermal \emph{P--T} profile of 3000~K, though they also tested the effects of varying the temperature to 2000~K and 4000~K. With the 3000~K model, they detected Ni~I at a significance of 5.01$\sigma$. Together, these results suggest that Ni~I is present in the atmosphere, and that the layer probed by Ni~I may be hotter than $\sim$2500~K. In the present work, we are unable to investigate the possibility of atmospheric Ni~I at a temperature of 3000~K. The results of this injection/recovery test are shown in Fig.~\ref{fig:NiI}.

\begin{figure*}
    \centering
    \includegraphics[]{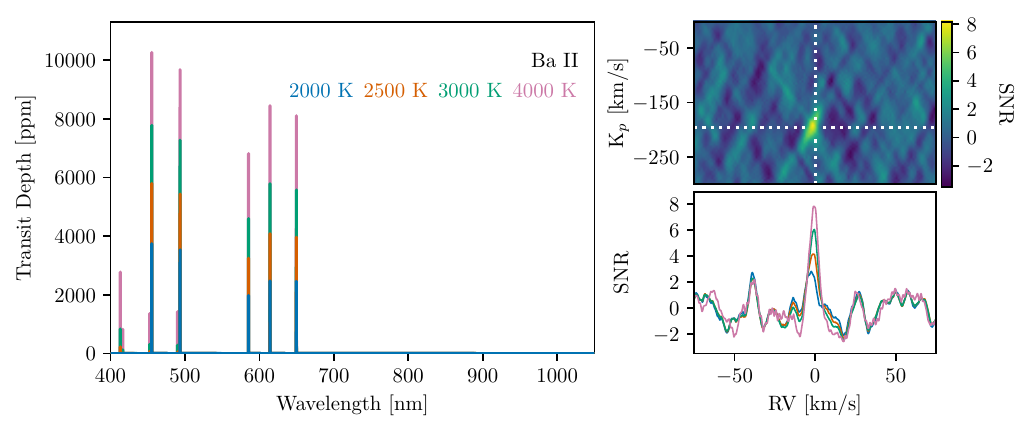}
    \caption{The model injection/recovery process for Ba~II. The plots are as described in the caption of Fig.~\ref{fig:NiI}, however in this case a 3000~K template is also displayed in green, and a 4000~K template (which resulted in the strongest recovered detection) is displayed in pink.}
    \label{fig:BaII}
\end{figure*}

In the case of Ba~II, which was detected by \cite{AS22} using the 2500~K Mantis Network template, only the 3000~K and 4000~K injected templates resulted in significant recovered detections of 6.1 and 8.1$\sigma$ respectively. On the other hand, we are unable to significantly detect the 2000~K or 2500~K templates. This could indicate that the atmospheric layer probed by Ba~II is cooler than 3000~K, and that our GRACES spectra are not sensitive to Ba~II at these cooler atmospheric temperatures (thus explaining why we did not detect the model that \cite{AS22} detects). The results of this injection/recovery test are shown in Fig.~\ref{fig:BaII}.

\begin{figure*}
    \centering
    \includegraphics{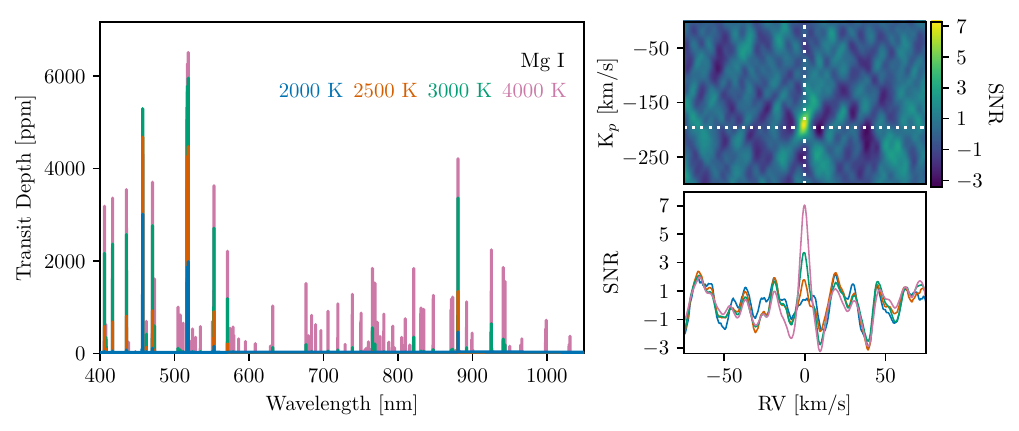}
    \caption{The model injection/recovery process for Mg~I. The plots are as described in the captions of Figs.~\ref{fig:NiI} and \ref{fig:BaII}.}
    \label{fig:MgI}
\end{figure*}

We are able to recover the 4000~K Mg~I model at a significance of 7.2$\sigma$, but do not recover any of the other Mg~I models. The results are shown in Fig.~\ref{fig:MgI}. This is not particularly surprising; as can be seen in Fig.~\ref{fig:MgI}, {the absorption lines present across the full GRACES wavelength range in the 4000~K template are in general stronger than those in the lower temperature templates.}
Mg~I was previously detected by \cite{Tabernero20}, \cite{Kesseli22}, and \cite{AS22}, all using ESPRESSO data. \cite{AS22} used the 2500~K Mantis Network template to make 4.4 and 9.5$\sigma$ detections (depending on the night), whereas \cite{Kesseli22} used a 3000~K model to detect Mg~I at a significance of 6.94$\sigma$ (though as with Ni~I, their detection strength doesn't vary significantly when 2000~K or 4000~K templates are used instead). On the other hand, \cite{Tabernero20} detected Mg~I through directly resolving absorption lines at $\sim$457 and $\sim$517nm; their detection strengths ranged from 2.8 to 7.5$\sigma$. Together, these results suggest that Mg~I would be detectable in GRACES spectra at a high ($\sim$4000~K) temperature, but that it is likely present at a lower temperature in the atmosphere of WASP-76b.

\begin{figure*}
    \centering
    \includegraphics{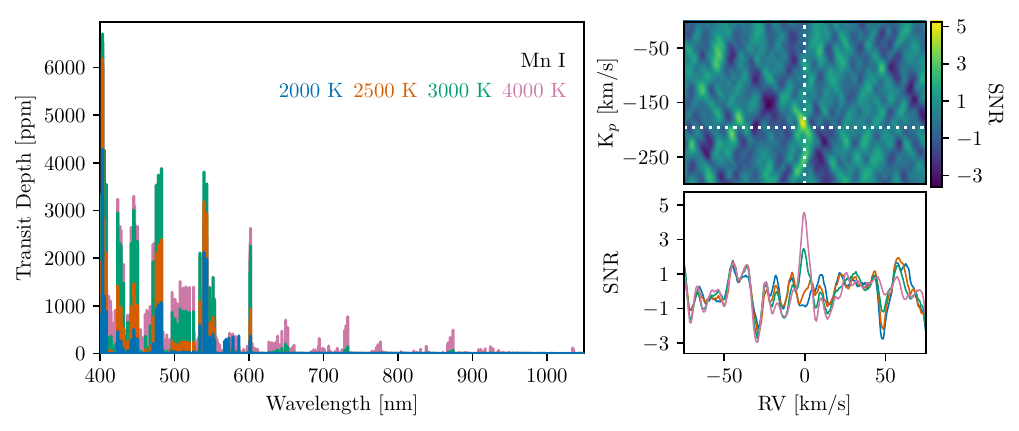}
    \caption{The model injection/recovery process for Mn~I. The plots are as described in the captions of Figs.~\ref{fig:NiI} and \ref{fig:BaII}.}
    \label{fig:MnI}
\end{figure*}

Likewise, we are only able to recover the 4000~K Mn~I model at a $>$5$\sigma$ significance. This is shown in Fig.~\ref{fig:MnI}. Mn~I was detected in ESPRESSO data by \cite{Tabernero20}, \cite{Kesseli22}, and \cite{AS22} (the latter using the 2500~K Mantis Network template); as with Mg~I, we conclude that Mn~I is likely present in the atmosphere at a temperature below 4000~K, which would not be detectable in our GRACES spectra.

\begin{figure*}
    \centering
    \includegraphics[]{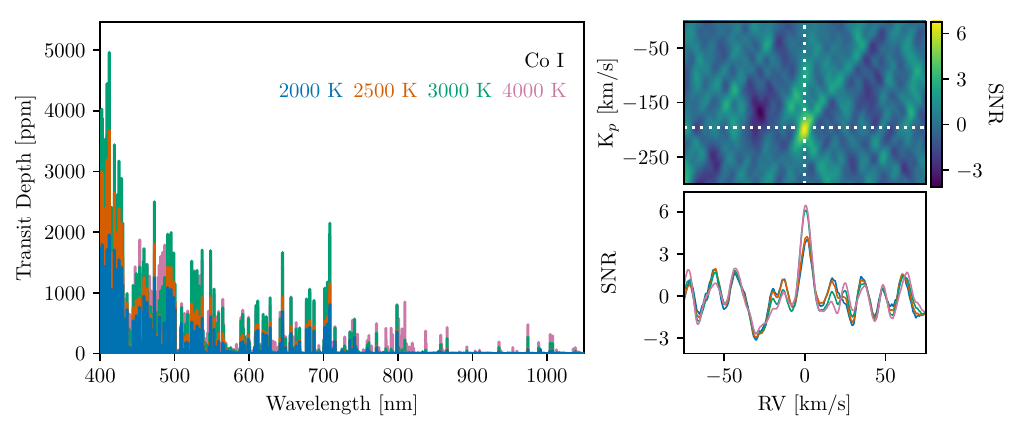}
    \caption{The model injection/recovery process for Co~I. The plots are as described in the captions of Figs.~\ref{fig:NiI} and \ref{fig:BaII}.}
    \label{fig:CoI}
\end{figure*}

We can recover the injected 3000~K and 4000~K Co~I templates at high significance (6.4 and 6.7$\sigma$ respectively) but are unable to recover the 2000~K or 2500~K templates at $>$5$\sigma$. Co~I was only tentatively detected by \cite{Kesseli22} at a significance of 4.03$\sigma$; our results indicate that the atmospheric layer probed by Co~I may be cooler than 3000~K, at a temperature to which our GRACES spectra are not sensitive. The results of this injection/recovery test are displayed in Fig.~\ref{fig:CoI}.

\begin{figure*}
    \centering
    \includegraphics[]{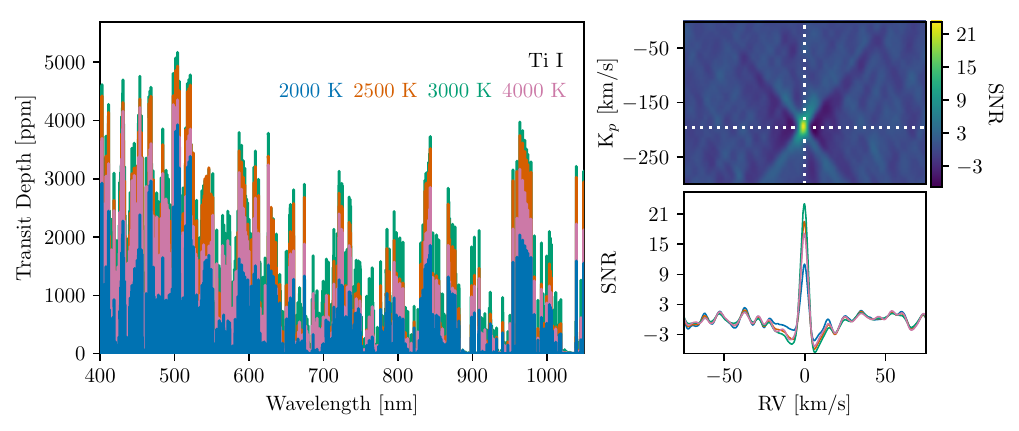}
    \caption{The model injection/recovery process for Ti~I. The plots are as described in the captions of Figs.~\ref{fig:NiI} and \ref{fig:BaII}.}
    \label{fig:TiI}
\end{figure*}

The Ti~I injections result in strong detections for all temperatures, with the 3000~K template recovered at the highest significance (23.1$\sigma$). As was discussed in \cite{Kesseli22}, Ti~I should be readily detectable in the atmosphere of WASP-76b, yet they were unable to detect it in their ESPRESSO observations. The fact that we are also unable to detect it in our original analysis indicates that Ti~I may be trapped in condensates via titanium cold-trapping \citep[e.g.,][]{Spiegel09, Parmentier13}. The results of this injection/recovery test are shown in Fig.~\ref{fig:TiI}.

\begin{figure*}
    \centering
    \includegraphics[]{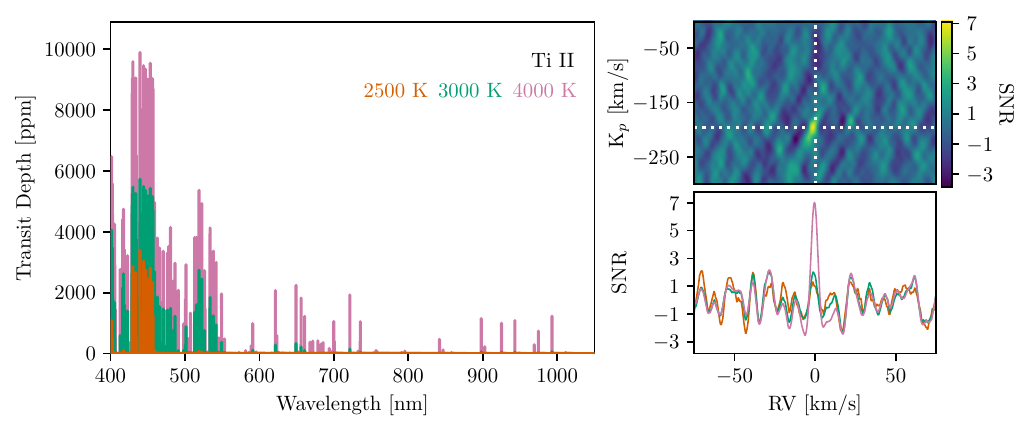}
    \caption{The model injection/recovery process for Ti~II. The plots are as described in the captions of Figs.~\ref{fig:NiI} and \ref{fig:BaII}.}
    \label{fig:TiII}
\end{figure*}

Ti~II, on the other hand, can only be recovered with the 4000~K template; this is shown in Fig.~\ref{fig:TiII}. Note that the 2000~K template contains no lines in the GRACES wavelength range. Combined with our Ti~I results, it's likely that any titanium present in the atmosphere is trapped in condensates.

\begin{figure*}
    \centering
    \includegraphics[]{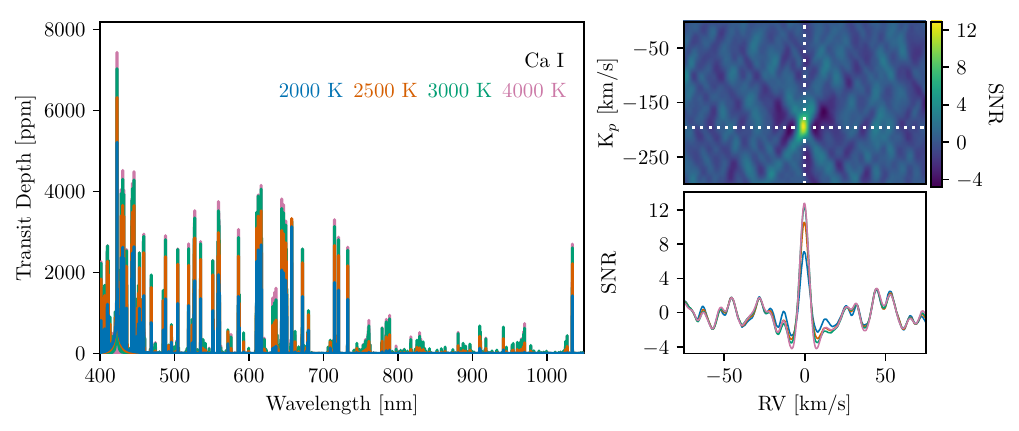}
    \caption{The model injection/recovery process for Ca~I. The plots are as described in the captions of Figs.~\ref{fig:NiI} and \ref{fig:BaII}.}
    \label{fig:CaI}
\end{figure*}

We are able to recover all of the Ca~I templates at a high ($>$5$\sigma$) significance. However given the strong Ca~II signals reported in the present work and \cite{Deibert21b}, it's likely that at the atmospheric layers probed by our observations, the majority of Ca~I has been ionized. This explains our non-detection of Ca~I, as well as the strong Ca~II signal detected in this work and \cite{Deibert21b}. The results of this injection/recovery test are displayed in Fig.~\ref{fig:CaI}.

\begin{figure*}
    \centering
    \includegraphics[]{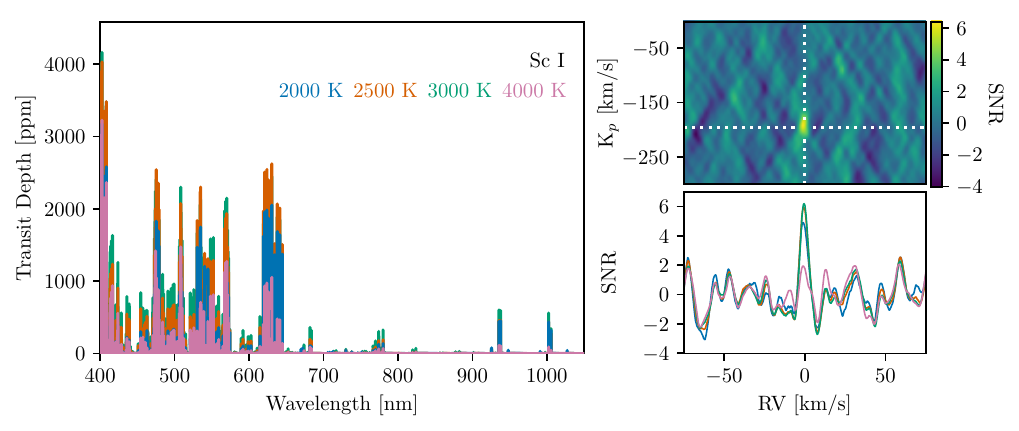}
    \caption{The model injection/recovery process for Sc~I. The plots are as described in the captions of Figs.~\ref{fig:NiI} and \ref{fig:BaII}.}
    \label{fig:ScI}
\end{figure*}

\begin{figure*}
    \centering
    \includegraphics[]{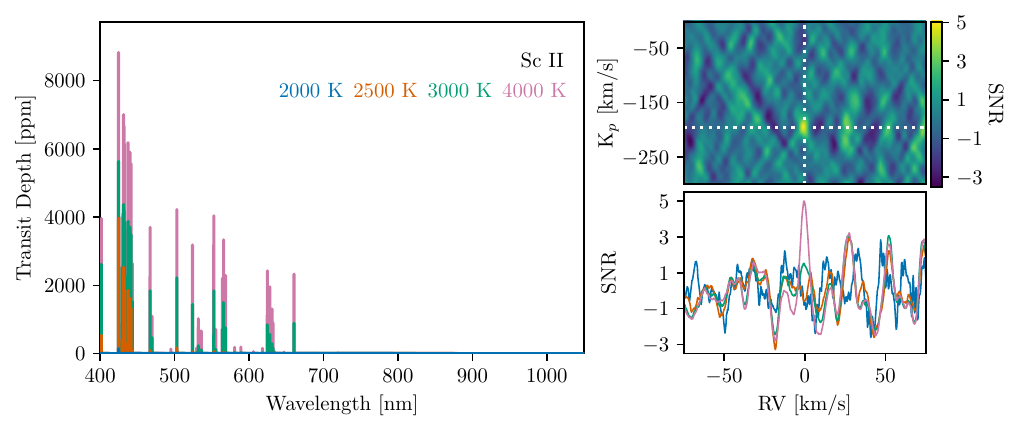}
    \caption{The model injection/recovery process for Sc~II. The plots are as described in the captions of Figs.~\ref{fig:NiI} and \ref{fig:BaII}.}
    \label{fig:ScII}
\end{figure*}

\begin{figure*}
    \centering
    \includegraphics[]{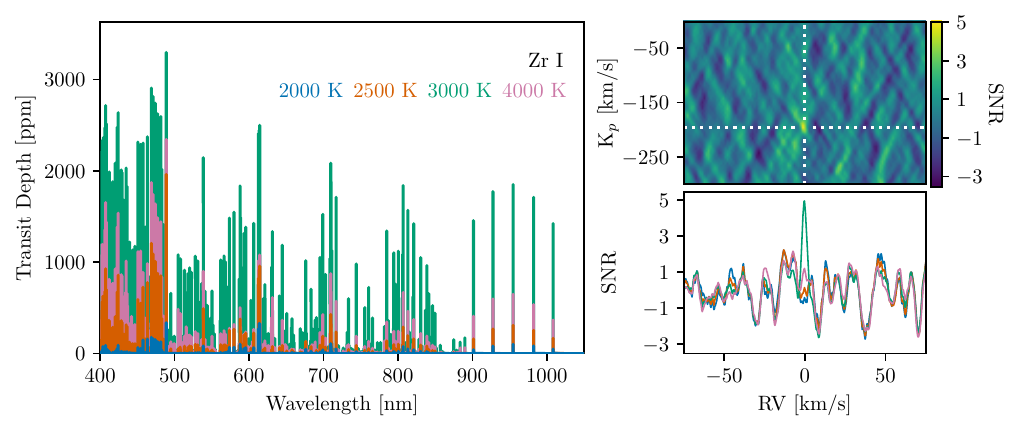}
    \caption{The model injection/recovery process for Zr~I. The plots are as described in the captions of Figs.~\ref{fig:NiI} and \ref{fig:BaII}.}
    \label{fig:ZrI}
\end{figure*}

We can recover the 2000~K, 2500~K, and 3000~K injected Sc~I templates, as well as the 3000~K Zr~I template, at high significances. Similarly, we can recover the 4000~K Sc~II model at a $>$5$\sigma$ significance, but do not recover the other models. Sc~I and Sc~II have not been detected in the atmosphere of WASP-76b; \cite{Kesseli22} show that both Sc~I and Sc~II should be marginally detectable (but were not detected) in ESPRESSO data. Combined with our model injection/recovery test, these results could indicate that Sc~I is not present in the atmosphere, or that it is ionized and present at a temperature $<$4000~K (as Sc~II has only a few lines in the optical at these cooler temperatures). 
Likewise, \cite{Kesseli22} could not detect Zr~I. They suggest that Zr~I may be ionized (and that Zr~II is not readily detectable in the atmosphere); our analysis also suggests that Zr~I could be present at a temperature cooler than $\sim$3000~K, which would not be detectable in our spectra. These results are shown in Fig.~\ref{fig:ScI} for Sc~I, Fig.~\ref{fig:ScII} for Sc~II, and \ref{fig:ZrI} for Zr~I.

\begin{figure*}
    \centering
    \includegraphics[]{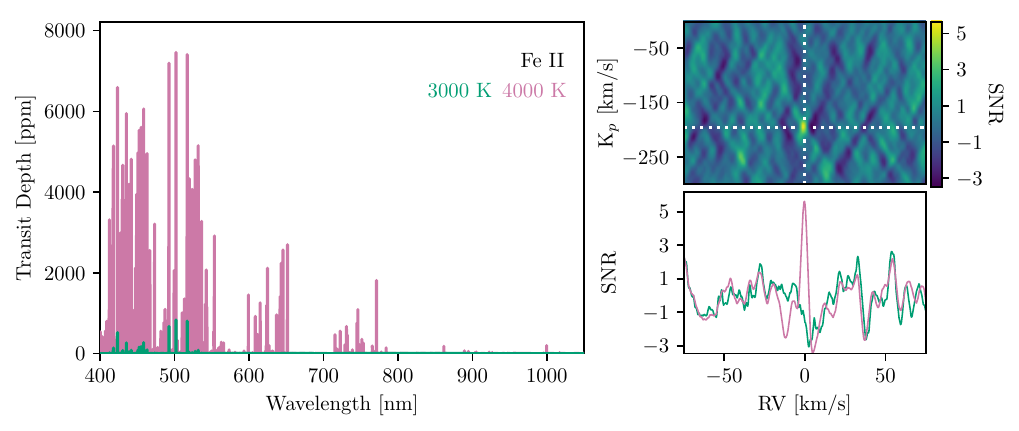}
    \caption{The model injection/recovery process for Fe~II. The plots are as described in the captions of Figs.~\ref{fig:NiI} and \ref{fig:BaII}.}
    \label{fig:FeII}
\end{figure*}

The 4000~K Fe~II template is recovered at a $>$5$\sigma$ significance, while we are unable to recover the 3000~K template. We do not include the 2000~K or 2500~K templates in our analysis as they do not contain any lines in the GRACES wavelength range. These results are shown in Fig.~\ref{fig:FeII}; they indicate that any Fe~II present in the regions probed is likely at a temperature lower than 4000~K, or that the Fe~I has mostly not ionized in the regions we're probing. 

\begin{figure*}
    \centering
    \includegraphics[]{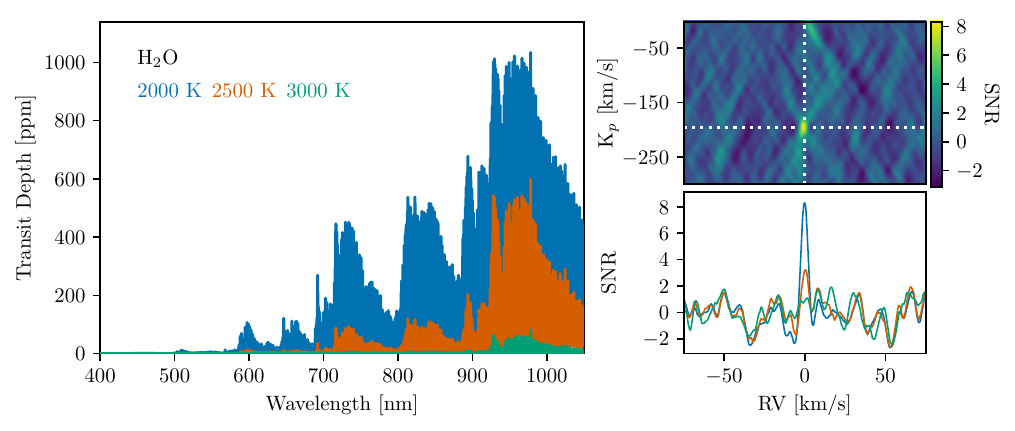}
    \caption{The model injection/recovery process for H${}_2$O. The plots are as described in the captions of Figs.~\ref{fig:NiI} and \ref{fig:BaII}.}
    \label{fig:H2O}
\end{figure*}

Finally, we are able to detect the injected 2000~K H${}_2$O template at a significance of 8.3$\sigma$, but are unable to detect the 2500~K or 3000~K templates. There are no lines present in the 4000~K template at the GRACES wavelength range. We note that \cite{SanchezLopez22} previously detected H${}_2$O at a significance of 5.5$\sigma$ in the atmosphere of WASP-76b, albeit with near-infrared spectra, using a pressure-temperature profile representative of the terminator region. Our results suggest that either H${}_2$O is present at an atmospheric layer hotter than 2000~K, which we are not sensitive to with our GRACES spectra, or that our telluric correction routine (see Section \ref{subsec:sysrem} and Appendix \ref{app:tellurics}) was unable to fully remove the telluric H${}_2$O absorption, which could have impacted our ability to detect water in the planet's atmosphere. The results of this injection/recovery test are shown in Fig.~\ref{fig:H2O}.

\subsection{Discussion}
We have recovered detections of Fe~I, Na~I, and Ca~II in the atmosphere of WASP-76b, and reported tentative detections of Li~I, K~I, Cr~I, and V~I. All of these species have previously been detected in WASP-76b's atmosphere; here we demonstrate that they are readily detectable with GRACES/Gemini North, and should be targets of interest in future ExoGemS analyses of other planets.

Interestingly, we also reported a number of non-detections, including for species which were previously detected with other instruments. There are several potential explanations for these discrepancies, and we investigated them further using model injection/recovery tests. 

In particular, Mg~I and Mn~I were previously detected by \cite{Tabernero20}, \cite{Kesseli22}, and \cite{AS22} using ESPRESSO, while \cite{Kesseli22} also detected Sr~II. We were unable to detect any of these species in our data; however, we were also unable to recover injected models of these species at temperatures cooler than 4000~K with our injection/recovery tests. This suggests that our observations are not sensitive to these species at cooler temperatures, but that they may well be present in WASP-76b's atmosphere (as indicated by \citealt{Tabernero20,Kesseli22,AS22}). These results are not particularly surprising given the fact that ESPRESSO extends to bluer wavelengths than GRACES, and all of these species exhibit strong absorption lines at the blue end of the optical regime. Furthermore, the higher spectral resolution of ESPRESSO (R $>$ 100,000) compared to GRACES (R $\sim$ 66,000) means that ESPRESSO can resolve a greater number of individual lines, likely boosting the instrument's detection capabilities.

In some cases, our non-detections can provide hints as to the atmospheric temperatures probed by various species. For example, while \cite{AS22} used a 2500~K template to detect Ba~II in their ESPRESSO observations, our model injection/recovery test demonstrated that Ba~II is not detectable in our GRACES data at a temperature of 2500~K, but would have been detected if it were present at a temperature of 3000~K. We find similar results for Co~I, suggesting that both of these species may be present at atmospheric layers with temperatures cooler than $\sim$3000~K. 

In other cases, our non-detections can help shed light on the atmospheric chemistry of WASP-76b. As in \cite{Kesseli22}, we find that Ca~I would have been readily detected if it were present in the atmosphere. Yet our inability to detect Ca~I, combined with our strong detection of Ca~II, indicates that the majority of calcium present in the region of the atmosphere we're probing has been ionized. This is consistent with other recent analyses of ultra-hot Jupiter atmospheres, in which a number of species are seen to be ionized as well \citep[e.g.,][]{AS22,Zhang22,Merritt21,Borsa21}.

Likewise, our non-detection of Ti~I in the atmosphere of WASP-76b provides further insight into its atmospheric chemistry. Through our model injection/recovery test, we showed that the high-resolution and broad wavelength coverage of GRACES make Ti~I readily detectable at a range of temperatures. Yet the fact that we are unable to detect Ti~I could indicate that titanium is depleted at the terminator region, perhaps because it has condensed on the cooler night-side of the planet and is trapped in condensates (for e.g., CaTiO${}_3$; i.e., a titanium cold-trap; \citealt{Spiegel09,Parmentier13}). Alternatively, titanium may condense at the terminator region only, yet still be present in the gas phase elsewhere in the atmosphere. Future work investigating the day-side emission spectrum of WASP-76b (as in e.g., \citealt{Hoeijmakers22}) could help shed light on whether titanium is depleted locally or globally.

In either case, this result adds to the growing body of work indicating that titanium is depleted in certain ultra-hot Jupiter transmission spectra \citep[e.g.,][]{Merritt21,Hoeijmakers22,Kesseli22}. Interestingly, Ti~I was recently detected in the transmission spectrum of WASP-189b \citep{Stangret22}, which has a slightly higher equilibrium temperature than WASP-76b (T${}_{\mathrm{eq}} \sim 2600$ K; \citealt{Anderson18}). TiO was also detected in WASP-189b's transmission spectrum by \citealt{Prinoth22}; likewise, Ti~II was detected in the transmission spectra of KELT-9b \citep[T${}_{\mathrm{eq}}$~$\sim$~4000~K;][]{Hoeijmakers19}. Together, these results suggest that Ti~I is depleted and potentially cold-trapped in ``cooler'' ultra-hot Jupiter atmospheres (i.e., T${}_{\mathrm{eq}} < \sim 2400$~K), while the hottest ultra-hot Jupiters do appear to exhibit titanium absorption in their atmospheres. Future analyses of ultra-hot Jupiters across a range of equilibrium temperatures will help shed light on these different atmospheric regimes. Our present study has demonstrated that GRACES is well-suited to investigating Ti~I depletion and cold-trapping in ultra-hot Jupiter atmospheres.

\subsubsection{Comparison Between Custom Models and Mantis Network Templates}
\label{subsec:callievmantis}
{While a full comparison between the results obtained with custom models and the Mantis Network templates for all species included here is beyond the scope of this work, we generated a Ca~II-only model using the methodology described in Section \ref{subsubsec:callie} in order to better compare the two methods. We are able to detect Ca~II with both models, as can be seen in Figs.~\ref{fig:callie-caii-correlation} and \ref{fig:CaII}. The 2000~K Mantis Network template is detected at a slightly higher significance of 5.3$\sigma$, whereas we detect the custom Ca~II model at a significance of 4.5$\sigma$.}

{The cross-correlations signals are broad for both models, which means the associated $K_p$ and $V_\mathrm{center}$ values have large errors. Nevertheless, we we find a Keplerian velocity of $K_p = 193^{+27}_{-17}$ km/s and wind speed of $V_{\mathrm{center}} = -7.1^{+5.1}_{-2.7}$ km/s for the Mantis template, while our custom model results in a Keplerian velocity of $K_p =  204^{+35}_{-46}$ km/s and a measured $V_\mathrm{center}$ of $-1.0{}^{+4.5}_{-4.8}$ km/s. While the wind speed values in particular are somewhat different from one another (though still within error), we note that the broad nature of these signals makes it difficult to accurately determine these parameters.}

{Overall, we conclude that both methods allow us to detect Ca~II, and that the Mantis Network templates can be useful in quickly exploring the effects of different temperatures in isothermal profiles on our detection capabilities.}

\section{Conclusion}
\label{sec:conclusion}
In this work, we presented the results of a cross-correlation analysis on spectra from one transit of the ultra-hot Jupiter WASP-76b. The observations were obtained with GRACES at the Gemini North telescope as part of the ongoing ExoGemS survey. Using both a custom one-dimensional transmission spectrum generated for WASP-76b's atmosphere as well as a grid of generic ultra-hot Jupiter atmospheric templates from the Mantis Network \citep{mantis}, we searched for absorption due to a suite of atomic and molecular features. We recover previous detections of Fe~I, Na~I, and Ca~II via cross-correlations with model templates, and report tentative detections as well as non-detections of a range of species, some of which had previously been detected with other instruments. These results allow us to assess the capabilities of GRACES compared to other high-resolution optical spectrographs while also validating our methodology for use in future ExoGemS analyses.

\section*{Acknowledgments}
%\begin{acknowledgments}
{We thank the referee for a thoughtful and constructive review which has greatly helped improve the quality of our work.}

We thank D.~Kitzmann, H.J.~Hoeijmakers, and collaborators for making their standard grid of masks and templates publicly available via the Mantis Network \citep{mantis}.

We also thank the staff of the Gemini North Observatory for their help in obtaining these observations. We extend a particular thank you to Teo Mocnik, the Contact Scientist for this program and observer for the data obtained in this work; Siyi Xu, the secondary Contact Scientist for this program; and Brittney Cooper, the telescope operator on duty during these observations.

{EKD acknowledges support from an NSERC Vanier Canada Graduate Scholarship and an NSERC Postdoctoral Fellowship.}

{RJ acknowledges support of a Rockefeller Foundation Bellagio Center residency.}

These observations were obtained through the Gemini Remote Access to CFHT ESPaDOnS Spectrograph (GRACES). ESPaDOnS is located at the Canada-France-Hawaii Telescope (CFHT), which is operated by the National Research Council of Canada, the Institut National des Sciences de l’Univers of the Centre National de la Recherche Scientifique of France, and the University of Hawai’i. ESPaDOnS is a collaborative project funded by France (CNRS, MENESR, OMP, LATT), Canada (NSERC), CFHT and ESA. ESPaDOnS was remotely controlled from the international Gemini Observatory, a program of NSF’s NOIRLab, which is managed by the Association of Universities for Research in Astronomy (AURA) under a cooperative agreement with the National Science Foundation on behalf of the Gemini partnership: the National Science Foundation (United States), the National Research Council (Canada), Agencia Nacional de Investigaci\'{o}n y Desarrollo (Chile), Ministerio de Ciencia, Tecnolog\'{i}a e Innovaci\'{o}n (Argentina), Minist\'{e}rio da Ci\^{e}ncia, Tecnologia, Inova\c{c}\~{o}es e Comunica\c{c}\~{o}es (Brazil), and Korea Astronomy and Space Science Institute (Republic of Korea).

This work was enabled by observations made from the Gemini North telescope, located within the Maunakea Science Reserve and adjacent to the summit of Maunakea. We are grateful for the privilege of observing the Universe from a place that is unique in both its astronomical quality and its cultural significance.
%\end{acknowledgments}

\facilities{Gemini:Gillett (GRACES), CFHT (ESPaDOnS)}

\software{
astropy \citep{astropy:2018},
barycorr \citep{barycorr}
batman \citep{batman},
IPython \citep{4160251},
Matplotlib \citep{Hunter:2007},
Numpy \citep{harris2020},
OPERA \citep{opera},
SciPy \citep{2020SciPy-NMeth}
}

\appendix

\section{\textsc{SysRem} Results}
\label{app:tellurics}
\begin{figure*}
    \centering
\includegraphics{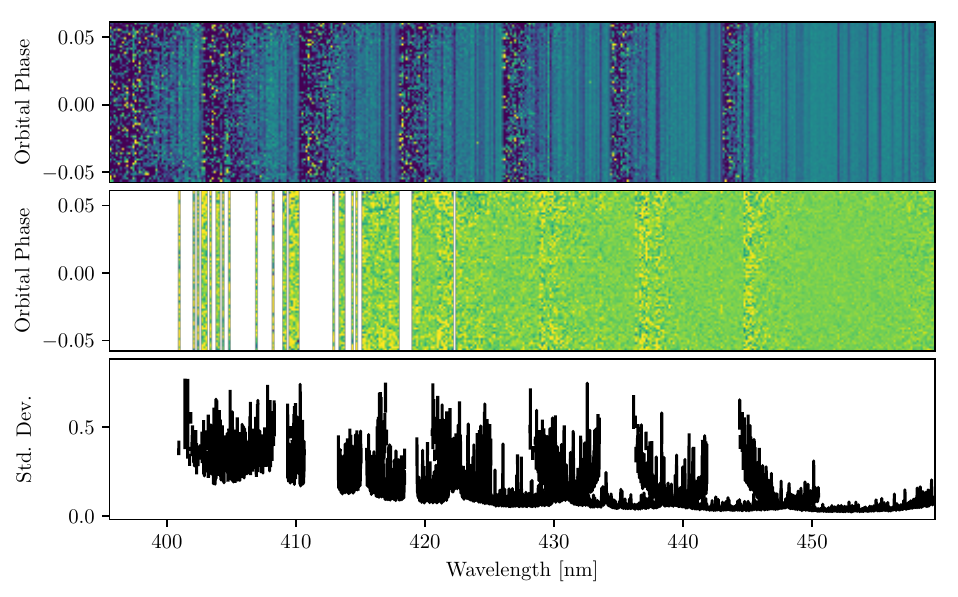}
    \caption{The results of applying the \textsc{SysRem} algorithm to 5 orders of the data at the blue end of the spectrum. \textit{Top}: The raw data after being extracted from the telescope and normalized. \textit{Middle}: The results of applying 7 iterations of the \textsc{SysRem} algorithm to the data. \textit{Bottom}: The standard deviation of the \textsc{SysRem}-reduced data along each wavelength channel. As described in Section \ref{subsec:detections}, all of these orders were excluded from our analysis.}
    \label{fig:sysr1}
\end{figure*}

\begin{figure*}
    \centering
    \includegraphics{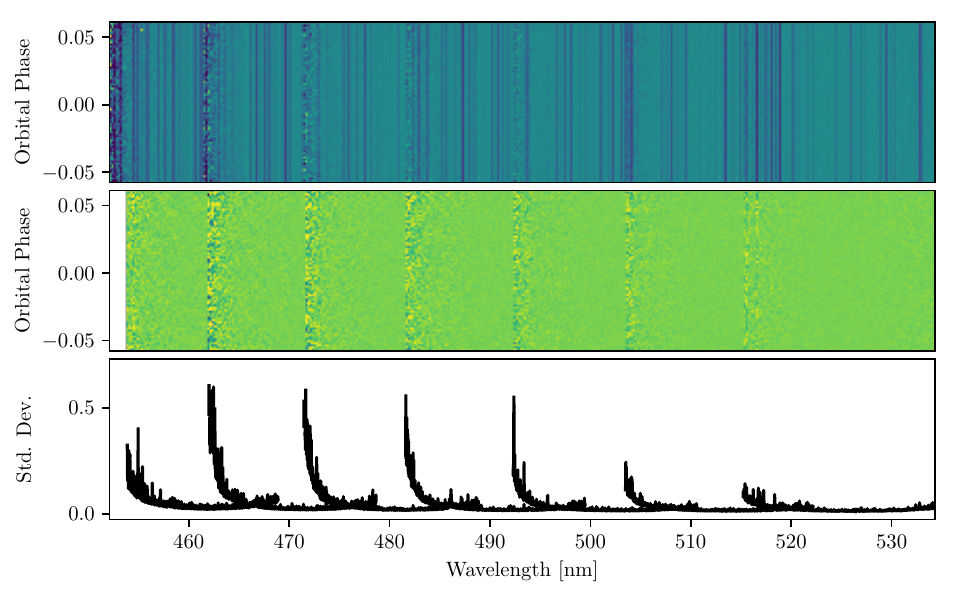}
    \caption{The results of applying the \textsc{SysRem} algorithm to 5 orders of the data. The panels are as described in the caption of Fig.~\ref{fig:sysr1}. One order at the blue end of this plot was excluded from our analysis. Note that the y-axis scale in the bottom plot differs from that in Fig.~\ref{fig:sysr1}.}
    \label{fig:sysr2}
\end{figure*}

\begin{figure*}
    \centering
    \includegraphics{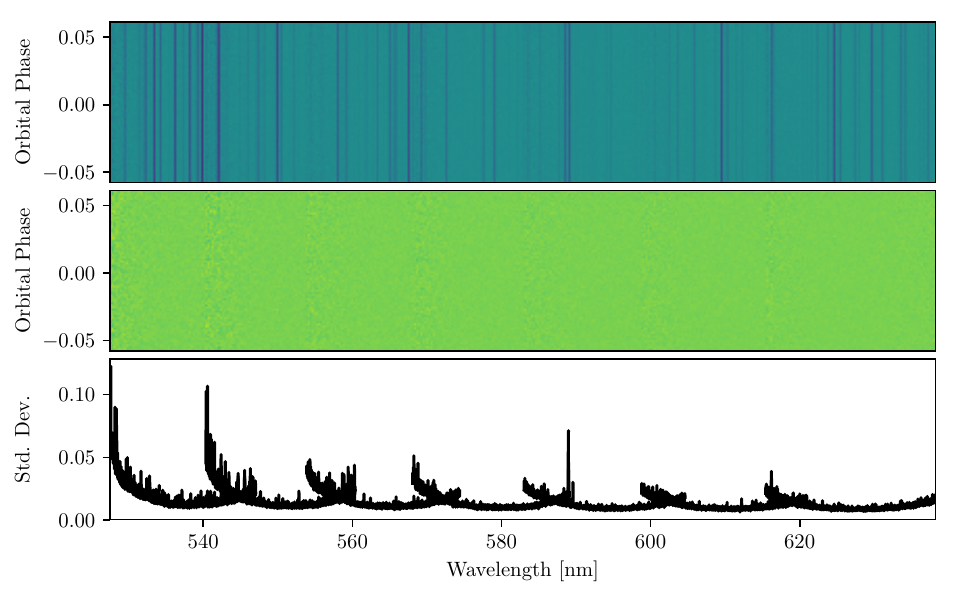}
    \caption{The results of applying the \textsc{SysRem} algorithm to 5 orders of the data. The panels are as described in the caption of Fig.~\ref{fig:sysr1}. Note that the y-axis scale in the bottom plot differs from that in Fig.~\ref{fig:sysr1}.}
    \label{fig:sysr3}
\end{figure*}

\begin{figure*}
    \centering
    \includegraphics{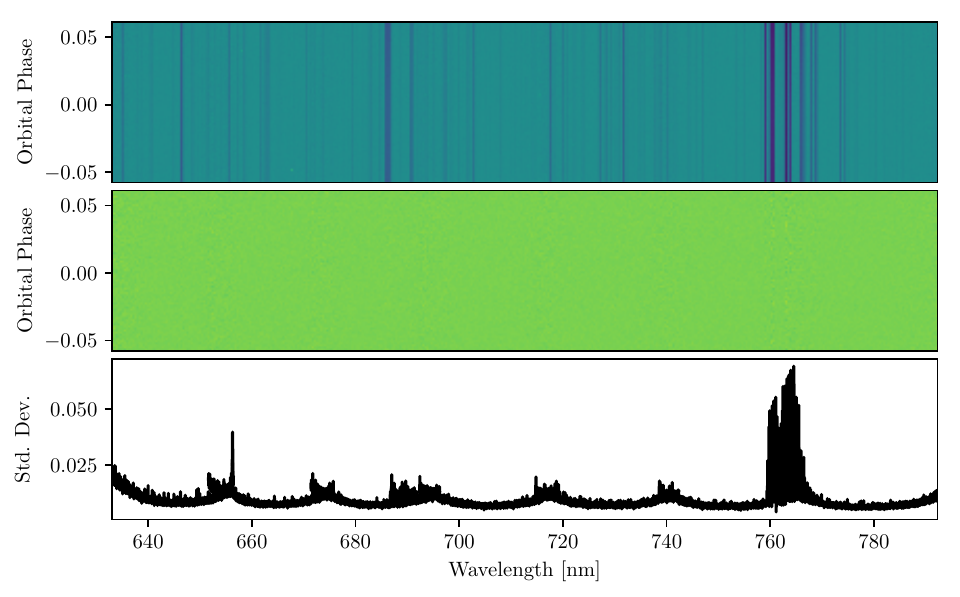}
    \caption{The results of applying the \textsc{SysRem} algorithm to 5 orders of the data. The panels are as described in the caption of Fig.~\ref{fig:sysr1}. Note that the y-axis scale in the bottom plot differs from that in Fig.~\ref{fig:sysr1}. Note as well the rise in the standard deviation around 760 nm due to greater telluric contamination.}
    \label{fig:sysr4}
\end{figure*}

\begin{figure*}
    \centering
    \includegraphics{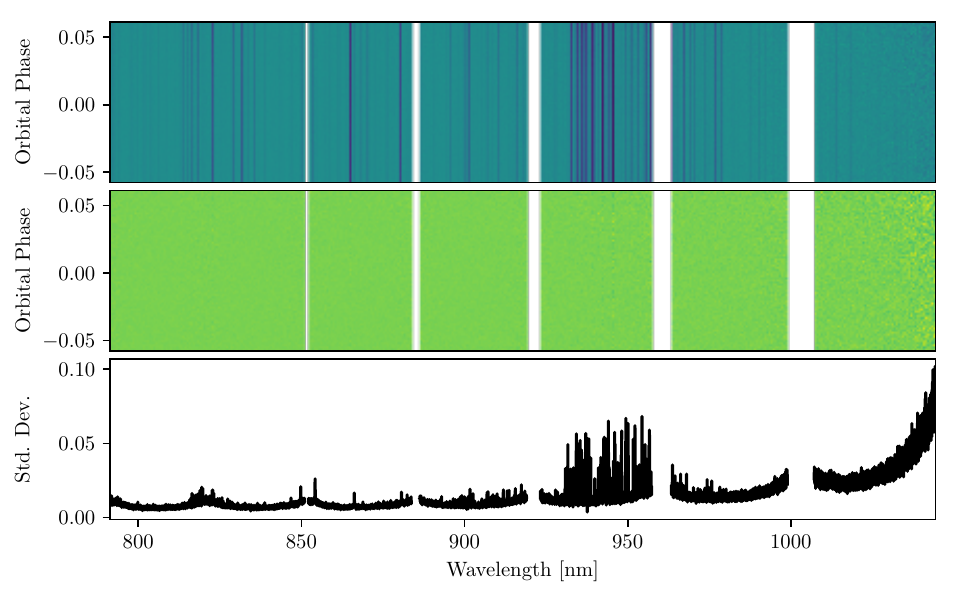}
    \caption{The results of applying the \textsc{SysRem} algorithm to 5 orders of the data at the red end of the spectrum. The panels are as described in the caption of Fig.~\ref{fig:sysr1}. Note that the y-axis scale in the bottom plot differs from that in Fig.~\ref{fig:sysr1}. Note as well the rise in the standard deviation around 940 nm due to greater telluric contamination.}
    \label{fig:sysr5}
\end{figure*}

{Figs.~\ref{fig:sysr1} to \ref{fig:sysr5} present the results of applying the \textsc{SysRem} algorithm to all orders of the data, as described in Section \ref{subsec:sysrem}. As can be seen in the figures, the standard deviation is much higher at the blue end of the spectrum (where the SNR of the raw data is lower, and the \textsc{SysRem} algorithm performed poorly). It rises again in regions that contain greater telluric contamination.}

\FloatBarrier

\section{Detection Significances for All Templates}
\label{app:sigs}
Table \ref{tab:detections-all} presents the detection significances for all templates corresponding to the species detected or tentatively detected in this work. We note that while the figures in Sections \ref{subsec:detections} and \ref{subsec:tentative} only display the template resulting in the highest SNR detection, the differences between different templates are in many cases small.

\begin{deluxetable}{ccccc}
\label{tab:detections-all}
\tablecaption{A summary of all detection significances and associated parameters for all temperatures for the detected ($>5\sigma$) and tentatively detected ($>3\sigma$) species. The columns are as described in the Table Notes of Table \ref{tab:detections-5}.}
\tablehead{%
    \colhead{Species}
    & \colhead{Temperature (K)}
    & \colhead{Significance ($\sigma$)}
    & \colhead{$K_p$ (km/s)}
    & \colhead{$V_\mathrm{center}$ (km/s)}
    }
\startdata %$^{+}_{-}$
 & 2000 & 6.6 & $183^{+7}_{-16}$ & $-4.6^{+1.2}_{-2.1}$ \\
Fe~I & 2500 & 6.9 & $182^{+8}_{-14}$ & $-4.4^{+0.9}_{-2.1}$ \\
 & 3000 & 6.8 & $181^{+9}_{-13}$ & $-4.4^{+1.5}_{-1.8}$ \\
 & 4000 & 6.0 & $183^{+10}_{-14}$ & $-3.8^{+1.8}_{-2.1}$ \\
 \hline
 & 2000 & 5.3 & $193^{+27}_{-17}$ & $-7.1^{+5.1}_{-2.7}$ \\
 Ca~II & 2500 & 4.0 & $217^{+14}_{-37}$ & $-3.5^{+4.5}_{-3.3}$ \\
 & 3000 & 2.3 & N/A & N/A \\
 & 4000 & 4.2 & $171^{+10}_{-12}$ & $-9.4^{+1.5}_{-3.3}$ \\
 \hline
  & 2000 & 5.0 & $201^{+11}_{-12}$ & $-2.5^{+1.5}_{-1.8}$ \\
  Na~I & 2500 & 4.2 & $200^{+14}_{-15}$ & $-2.5 \pm 1.8$ \\
  & 3000 & 4.7 & $203^{+10}_{-12}$ & $-2.3^{+1.5}_{-1.8}$ \\
  & 4000 & 4.1 & $201^{+11}_{-13}$ & $-2.8^{+1.8}_{-1.5}$ \\
  \hline
  \hline
   & 2000 & 3.9 & $222^{+18}_{-13}$ & $0.4 \pm 2.4$ \\
   Li~I & 2500 & 4.2 & $221^{+17}_{-14}$ & $0.7 \pm 2.4$ \\
     & 3000 & 4.0 & $221^{+19}_{-13}$ & $0.7^{+2.4}_{-2.1}$ \\
     & 4000 & 3.8 & $225^{+18}_{-16}$ & $1.0^{+2.7}_{-2.4}$ \\
     \hline
     & 2000 & 3.9 & $244^{+9}_{-40}$ & $-4.1 \pm 1.5$ \\
     K~I & 2500 & 4.3 & $211^{+35}_{-8}$ & $-5.3^{+1.2}_{-1.5}$ \\
     & 3000 & 4.4 & $209^{+29}_{-6}$ & $-5.6 \pm 1.2$ \\
     & 4000 & 4.3 & $209^{+28}_{-7}$ & $-5.6 \pm 1.2$ \\
     \hline
     & 2000 & 0.7 & N/A & N/A \\
     Cr~I & 2500 & 4.0 & $209^{+8}_{-10}$ & $3.5^{+1.5}_{-1.2}$ \\
     & 3000 & 4.0 & $208^{+8}_{-11}$ & $3.2^{+1.8}_{-1.2}$ \\
     & 4000 & 3.8 & $206 \pm 10$ & $3.2^{+1.5}_{-13.8}$ \\
     \hline
     & 2000 & 3.6 & $142^{+81}_{-17}$ & $-8.0^{+2.1}_{-1.8}$ \\
     V~I & 2500 & 4.0 & $143^{+78}_{-18}$ & $-7.7^{+1.8}_{-2.1}$ \\
     & 3000 & 4.2 & $140^{+14}_{-16}$ & $-7.7^{+1.8}_{-2.1}$ \\
     & 4000 & 4.6 & $140^{+13}_{-16}$ & $-7.4^{+1.8}_{-2.1}$ \\
\enddata
\tablecomments{For templates which were neither detected nor tentatively detected, we cannot extract a $K_p$ or $V_\mathrm{center}$ value; thus, we have marked those values as N/A.}
\end{deluxetable}

\section{Cross-Correlation Maps for Non-Detections}
\label{app:nondetec-5}
\begin{figure*}
\centering
\includegraphics[width=\textwidth]{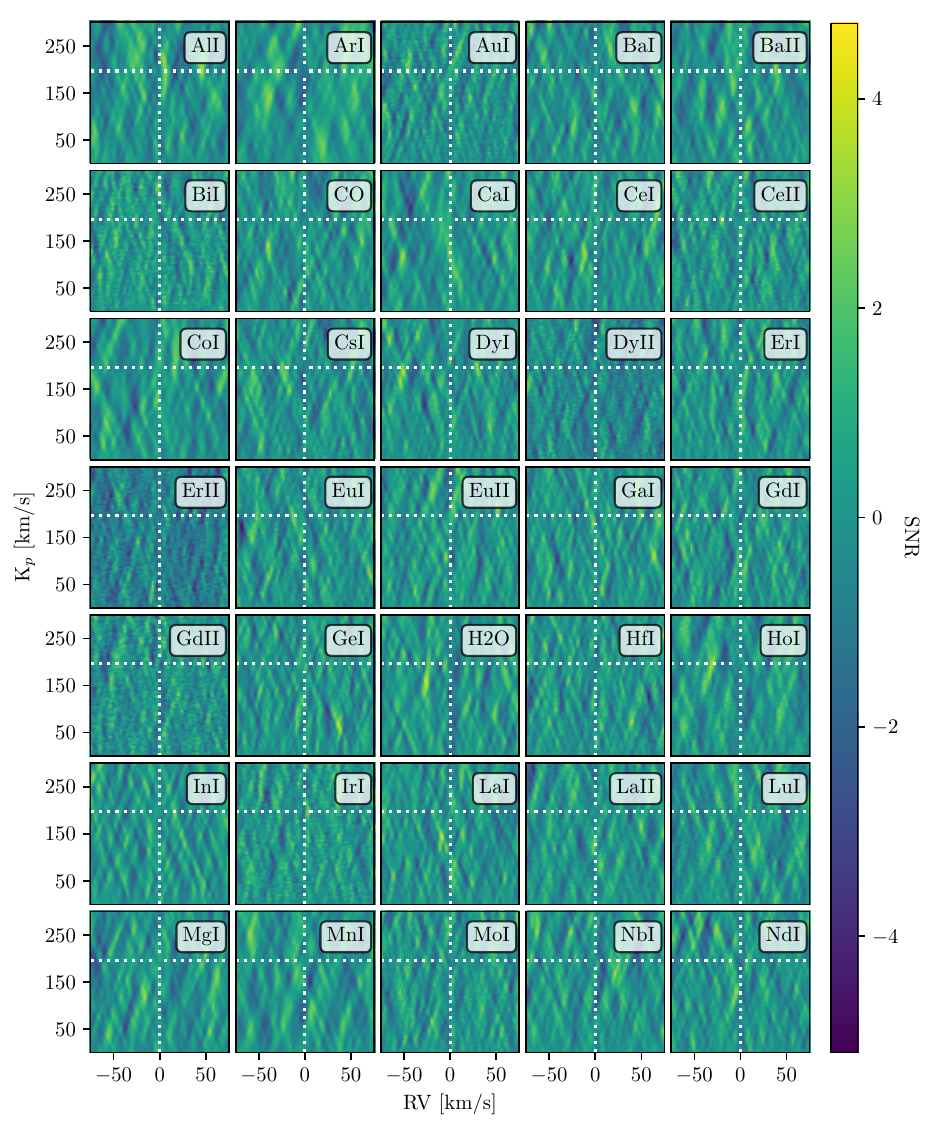}
\caption{The 2D $K_p$-RV maps for species which were not detected in this analysis. All plots were made using the 2000~K template; higher-temperature templates also resulted in non-detections. The species is indicated in the top right corner. The white lines indicate the expected $K_p$ and RV of the signal.}
\label{fig:non-part1}
\end{figure*}

\begin{figure*}
\centering
\includegraphics[width=\textwidth]{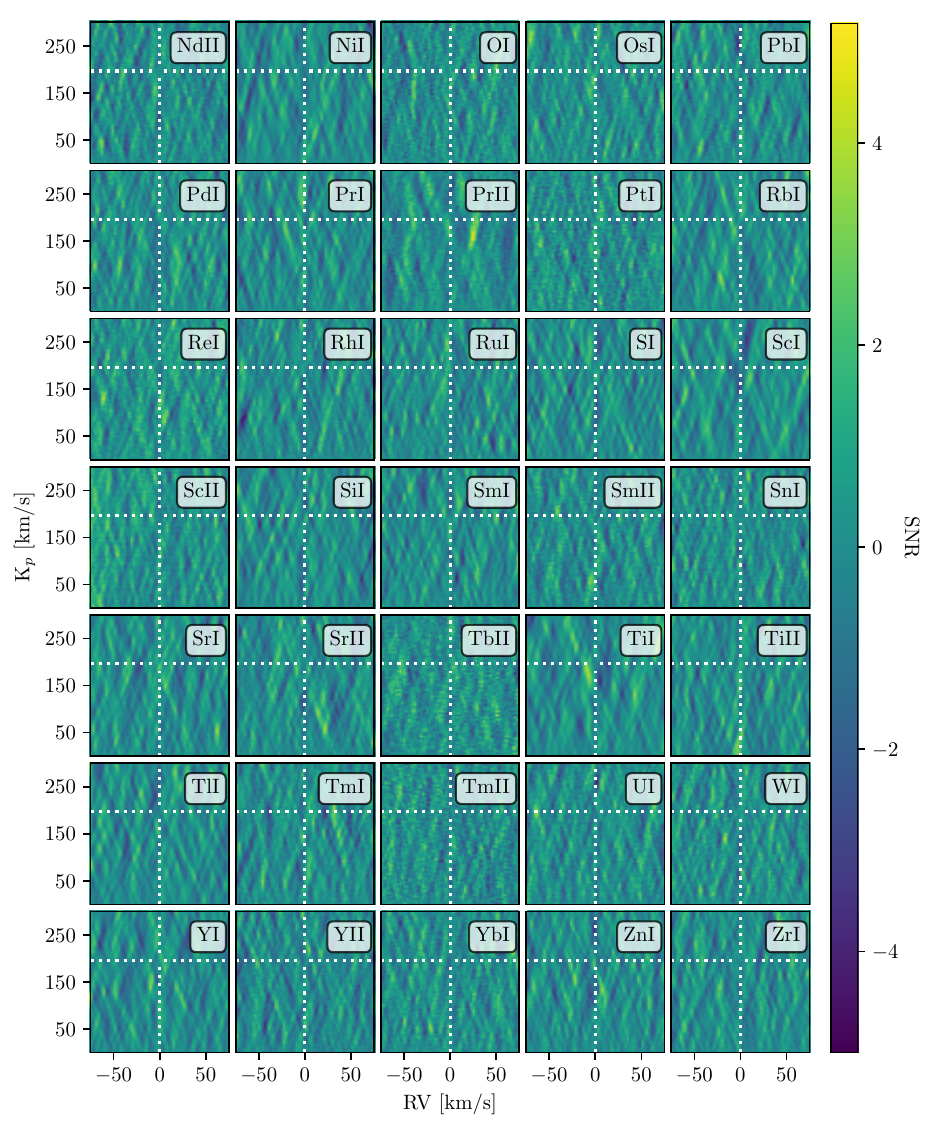}
\caption{The 2D $K_p$-RV maps for species which were not detected in this analysis. The figures are as described in the caption of Fig.~\ref{fig:non-part1}.}
\label{fig:non-part2}
\end{figure*}

Figs.~\ref{fig:non-part1} and \ref{fig:non-part2} present the 2D $K_p$-RV maps for the species which were not detected in this work. For the sake of brevity, we have only included the 2000~K models. We define a non-detection as being under 3$\sigma$.

\FloatBarrier

\bibliography{references}{}
\bibliographystyle{aasjournal}

\end{document}